\documentclass[prl,twocolumn,nofootinbib]{revtex4-1}
\usepackage{graphicx, epsfig}
\usepackage[dvipsnames]{xcolor}
\usepackage{hyperref}
\usepackage{mathrsfs}
\usepackage{amsmath}
\usepackage{enumerate}
\usepackage{bm}

\usepackage{float}
\usepackage{ulem}

\newcommand{\be}{\begin{equation}}
\newcommand{\ee}{\end{equation}}
\newcommand{\bea}{\begin{eqnarray}}
\newcommand{\eea}{\end{eqnarray}}
\newcommand{\ba}{\begin{eqnarray}}
\newcommand{\ea}{\end{eqnarray}}

\newcommand{\gapp}{\mathrel{\raise.3ex\hbox{$>$}\mkern-14mu
              \lower0.6ex\hbox{$\sim$}}}
\newcommand{\lapp}{\mathrel{\raise.3ex\hbox{$<$}\mkern-14mu
              \lower0.6ex\hbox{$\sim$}}}

\begin{document}
\title{ Probing the electroweak symmetry breaking history with Gravitational waves }

\author{Zizhuo Zhao}
\affiliation{Department of Physics, Chongqing University, Chongqing 401331, P. R. China}
\author{Yuefeng Di}
\affiliation{Department of Physics, Chongqing University, Chongqing 401331, P. R. China}

\author{Ligong Bian }\email{lgbycl@cqu.edu.cn}
\affiliation{Department of Physics and Chongqing Key Laboratory for Strongly Coupled Physics, Chongqing University, Chongqing 401331, P. R. China}
\affiliation{ Center for High Energy Physics, Peking University, Beijing 100871, China}

\author{Rong-Gen Cai}\email{cairg@itp.ac.cn}
\affiliation{CAS Key Laboratory of Theoretical Physics, Institute of Theoretical Physics, Chinese Academy of Sciences, P.O. Box 2735, Beijing 100190, China}
\affiliation{School of Physical Sciences, University of Chinese Academy of Sciences, No. 19A Yuquan Road, Beijing 100049, China}
\affiliation{School of Fundamental Physics and Mathematical Sciences, Hangzhou Institute for Advanced Study, University of Chinese Academy of Sciences, Hangzhou 310024, China}

\begin{abstract}

We perform a three dimensional lattice simulation of the electroweak symmetry breaking process through a two-step phase transition, where one of the two steps is a first order phase transition. Our results show that: 1) when the electroweak symmetry breaking is driven by the beyond Standard Model sector around $\sim \mathcal{O}(10^{2-3})$ GeV, the gravitational wave spectra produced from the phase transitions are of broken power-law double-peak shapes; 2) when the electroweak symmetry breaking is induced by a first-order phase transition of a high-scale global U(1) theory, cosmic strings can form and then disappear through particle radiation, and the yielded gravitational wave spectra are of plateau shapes. 
The two scenarios can be distinguished through probing gravitational wave spectra. Our study suggests that  the stochastic gravitational waves provide an alternative way to probe the beyond Standard Model sector relevant to the electroweak symmetry breaking pattern in the early Universe.  

\end{abstract}

\maketitle

\noindent{\it \bfseries  Introduction:}
While the phase transition (PT) pattern in the Standard Model (SM) of particle physics is {\it cross-over}~\cite{DOnofrio:2014rug}, a first-order PT is a general prediction in many new physics models beyond the SM~\cite{Mazumdar:2018dfl,Caldwell:2022qsj}.
The first-order PTs can produce 
stochastic gravitational wave (GW) backgrounds, which are detectable by  LIGO and Virgo~\cite{Abbott:2016blz}, Laser Interferometer Space Antenna (LISA)~\cite{Audley:2017drz}, Taiji~\cite{Guo:2018npi}, TianQin~\cite{Luo:2015ght},  Big Bang Observer (BBO)~\cite{Corbin:2005ny}, and DECi-hertz Interferometer Gravitational wave Observatory (DECIGO)~\cite{Yagi:2011wg}, etc. Recently, the constraints on 
new physics admitting low-scale and high-scale first-order PTs are placed by PPTA~\cite{Xue:2021gyq} (and NANOGraV~\cite{NANOGrav:2021flc}) and LIGO-Virgo~\cite{Romero:2021kby,Jiang:2022mzt}. Therefore, the stochastic GW background searches open a new astronomy window to probe new physics beyond the SM~\cite{Mazumdar:2018dfl,Caprini:2015zlo,Caprini:2019egz}.

The Early Universe may settle down to the electroweak vacuum through multi-step phase transitions. 
Firstly, the electroweak symmetry breaking (EWSB) process can occur through two-step electroweak PTs with one or two steps being first-order, where the Baryon Asymmetry of the Universe can be explained through electroweak baryogenesis~\cite{Kuzmin:1985mm,Shaposhnikov:1986jp,Shaposhnikov:1987tw,Morrissey:2012db,Patel:2012pi,Blinov:2015sna,Inoue:2015pza,Ramsey-Musolf:2017tgh,Xie:2020bkl,Jiang:2015cwa}, the dark matter can be accommodated together with strong GW signals to be probed by LISA and other GW detectors~\cite{Jiang:2015cwa,Bian:2018mkl,Bian:2018bxr,Baker:2016xzo,Baker:2017zwx,Chao:2017vrq}. Secondly, the EWSB may occur through dimensional transmutation after hidden sectors undergo first-order PT with the detectable GWs ~\cite{Schwaller:2015tja,Jaeckel:2016jlh,Croon:2018erz,Breitbach:2018ddu,Fairbairn:2019xog,Baldes:2017rcu,Tsumura:2017knk,Aoki:2017aws,Croon:2018new,Baldes:2018emh}, as in many classically conformal theories motivated for the gauge hierarchy problem~\cite{Foot:2007iy,Iso:2009nw,Englert:2013gz,Farzinnia:2013pga,Hur:2011sv,Chang:2007ki,Iso:2009ss}. Therein, interactions between the SM Higgs and hidden sectors may keep in thermal equilibrium or not in the early Universe. Thirdly, EWSB might occur after a global U(1) symmetry breaking whose goldstone can be the axion for the solution of the strong CP problem~\cite{Vilenkin:1986ku, Davis:1986xc,Harari:1987ht,Hagmann:1990mj,Battye:1993jv,Battye:1994au,Yamaguchi:1998gx,Hagmann:2000ja}, where the cosmic strings can form during the spontaneous symmetry breaking process~\cite{Kibble:1976sj,Hindmarsh:1994re}.  GWs from first-order PT in axion-like particle models might be able to probed by LIGO~\cite{Dev:2019njv,VonHarling:2019rgb,Ghoshal:2020vud,DelleRose:2019pgi,Romero:2021kby}. 
In this Letter, we numerically study the dynamical EWSB driven by the above three classes of two-step PTs in the early Universe, and investigate the feature of associated  stochastic GW backgrounds.
Previous lattice simulations of the GWs produced during the first-order PT process usually adopt a single scalar field one-step PT model~\cite{Giblin:2014qia,Hindmarsh:2013xza,Cutting:2019zws,Hindmarsh:2015qta,Hindmarsh:2017gnf,Cutting:2018tjt,Cutting:2020nla,Pol:2019yex}. To the best of our knowledge, the three-dimensional lattice simulation of first-order PT with two scalars haven't been studied before.
In this Letter, we investigate EWSB occuring through three classes of two-step PTs by performing three-dimensional lattice simulation considering dynamics of Higgs and beyond SM scalar. We study the produced GWs spectra of three types of  two-step phase transitions with the first- or second- step being first-order PT. We also investigate the formation and evolution of cosmic strings and their effect on the produced GWs during the first-order PT.  

\begin{figure}[!htp]
\begin{center}
\includegraphics[width=0.2\textwidth]{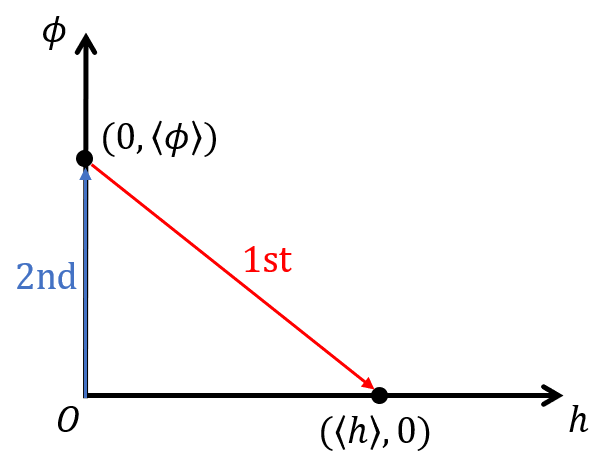}
\includegraphics[width=0.2\textwidth]{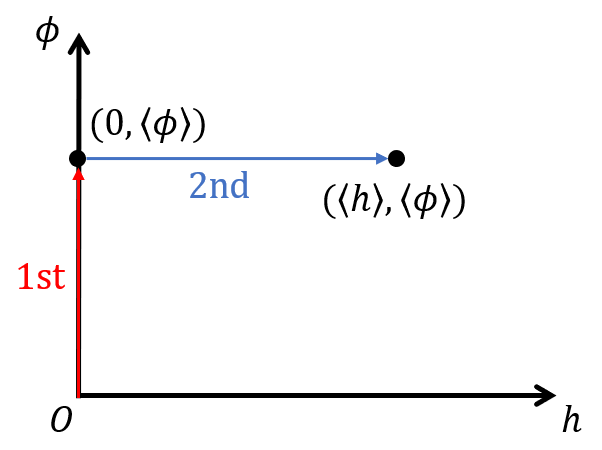}
\caption{PT patterns for type-a (left), type-b and type-c classes (right).  } \label{ptpatterns}
\end{center}
\end{figure}

\noindent{\it \bfseries PT models: }
\label{sec:model}
We introduce three classes of PT models for the study of dynamical EWSB and GWs production. 
The first class of PT model (type-a) is motivated by dark matter and baryogenesis~\cite{Bian:2018bxr,Bian:2018mkl,Baker:2016xzo,Baker:2017zwx,Chao:2017vrq,Patel:2012pi,Blinov:2015sna,Inoue:2015pza,Ramsey-Musolf:2017tgh,Xie:2020bkl,Jiang:2015cwa}, with the first-step being a {\it cross-over}  and the second-step being a first-order PT to produce GWs. Since the first-step second-order PT is known to yield null GWs, we focus on the second step first-order PT with the vacuum transiting
from the dark vacuum ($(0,\langle \phi \rangle)$) to the electroweak vacuum ($ (\langle h \rangle,0)$), see left plot of Fig.~\ref{ptpatterns}. 
We adopt the gauge invariant form of the thermal effective potential 
\begin{eqnarray}
V_a(\phi,h,T)&=&\frac{1}{2}(\mu_\phi^2+c_\phi T^2)\phi^2+\frac{1}{2}\lambda_{h\phi}h^2\phi^2+\frac{1}{4}\lambda_\phi\phi^4\nonumber\\
&+&\frac{1}{2}(-\mu_h^2+c_h T^2)h^2+\frac{1}{4}\lambda_h h^4\;,
\end{eqnarray}
with $c_\phi=\lambda_\phi/4+\lambda_{h\phi}/3\;, c_h=(2 m_W^2+m_Z^2+2 m_t^2)/(4 v^2)+\lambda_h/2+\lambda_{h\phi}/12\;$. 
Here, $m_{W,Z,t}$ are masses of W (Z) bosons and top quark, the $v$ is VEV of the SM Higgs. In this model, we consider the dark sector can keep in thermal equilibrium with the SM 
with a moderate Higgs coupling $\lambda_{h\phi}\sim\mathcal{O}(10^{-1})$.

The second class of PT models (type-b) are motived by the solution of gauge hierarchy problem~\cite{Foot:2007iy,Iso:2009nw,Englert:2013gz,Farzinnia:2013pga,Hur:2011sv,Chang:2007ki,Iso:2009ss}, where the underlying theory inducing EWSB is a classical conformal theory (CCT) admitting first-order PT~\cite{Jinno:2016knw,Marzola:2017jzl,Iso:2017uuu,Lewicki:2020jiv}.  
After the first-step first-order PT occurs in the CCT ($(0,0)\to  (0, \langle \phi\rangle)$) with the production of GWs, the vacuum transits from the symmetric phase to CCT's vacuum, the dimensional transmutation process in the second-step PT would induce EWSB ($(0, \langle \phi\rangle)\to (\langle h\rangle,\langle \phi\rangle)$), see right plot of Fig.~\ref{ptpatterns}.
The thermal potential contribution from CCT takes the form of, 
\begin{eqnarray}\label{vb1}
  V_{cct}(\phi,T)&=&a \phi^4(\log[|\phi|^2/v_\phi^2]-1/4)+ b T^2 |\phi| ^2\;, 
    \end{eqnarray}
and the relevant thermal potential for dynamical dimensional transmutation process is 
  \begin{eqnarray}\label{vb2}
  V_{dt}(\phi,h,T)&=&\frac{1}{2}c^\prime_hT^2 h^2+\frac{1}{4}\lambda_h h^4-\frac{\lambda_p}{4} h^2\phi^2\;,
\end{eqnarray}
with $c^\prime_h=(2 m_W^2+m_Z^2+2 m_t^2)/(4 v^2)+\lambda_h/2+\lambda_p/24\;$. The Higgs-portal coupling $\lambda_p$ controls the splitting between the electroweak scale and the high scale  CCT (characterized by the $v_\phi$). 

Motivated by the global strings intimately connected with strong CP problem~\cite{Vilenkin:1986ku, Davis:1986xc,Harari:1987ht,Hagmann:1990mj,Battye:1993jv,Battye:1994au,Yamaguchi:1998gx,Hagmann:2000ja} and the axion dark matter physics~\cite{Caldwell:2022qsj,Buschmann:2019icd,Gorghetto:2018myk,Figueroa:2020lvo,Gorghetto:2021fsn,Chang:2021afa}, we consider the third PT model where the EWSB is induced by spontaneously symmetry breaking of a high scale global U(1) theory after a first-order PT (namely type-c)\footnote{For early studies on the cosmic defects that might be formed due to ``geodesic rule" when vacuum bubbles collide with each other, see Refs.~\cite{Borrill:1995gu,Digal:1997ip,Copeland:1996jz, Digal:1997gc,Ferrera:1995ef,Ferrera:1997xg,Ferrera:1996hu,Digal:1997gc,Copeland:1999ua,Davis:1999ii,Lilley:2001df}. }. We study the GWs associated with formation and decay of global strings during this type PT. 
We consider the phase transition pattern of type-c PT to be similar with type-b PT, with the vacuum transiting from the symmetric phase to the U(1) vacuum, and then to the electroweak vacuum as shown in the right plot of Fig.~\ref{ptpatterns}. Here, we take the same thermal potential form as in Eqs.(\ref{vb1},\ref{vb2}) with the real scalar $\phi$ replaced by a complex scalar ($\Phi$) persevering a global U(1) symmetry.

In the simulations of the three type PTs,  we safely neglect the effect of cosmic expansion because the PTs under consideration are fast, and therefore take the equation of motions for the real scalar fields as:
\begin{equation}\label{scaeq}
\phi_{i}^{\prime\prime} -\nabla^2\phi_{i} + \frac{dV}{d\phi_{i}}=0\text,
\end{equation}
with the thermal potential $V$ given above. $\phi_{i}=h,\phi$ for type-a and type-b PTs, for type-c PT we include also the equation of motion for the Higgs field except for that of the complex scalar $\Phi$ with $\phi_i=h,\phi_{1,2}$.  
At the beginning, we have the initial conditions $\phi_{i}=\dot\phi_{i}=0$. 
 For type-a PT, when bubbles start to nucleate, we have the bubble profiles for the second-step first-order PT, 
\begin{eqnarray} 
\label{eq:Phiini}
h(t=0,{r}) =  \eta_h/2\left [ 1-\tanh \left ( \frac{r-R_0}{L_w}\right ) \right ]\;,\\
\phi (t=0,{ r}) =  \eta_\phi/2\left [ 1+\tanh \left ( \frac{r-R_0}{L_w}\right ) \right ]\;,
\end{eqnarray}
where $R_0$ is the initial bubble radius and $L_w$ is  the thickness of the critical bubble wall. Here, $\eta_{h,\phi}$ are vacuum expectation values (VEVs) of dark vacuum and electroweak vacuum when PT occurs.
For the type-b PT, the bubble profiles admit the form of $\phi (t=0,{r}) =  \eta_\phi/2\left [ 1- \tanh ( r-R_0)/L_w \right ]$ with $\eta_\phi$ being the VEV of $\phi$ when the CCT vacuum bubbles nucleate, and $\langle h\rangle =\sqrt{(\lambda_p \eta^2-2 c^\prime_hT^2)/(2 \lambda_h)}$ inside these bubbles. For the type-c PT, we consider the two real scalar fields of U(1), $\Phi=(\phi_1,\phi_2)$, $\phi_1=\phi(t=0,{\bm r})\cos\theta/2 $ and $\phi_2=\phi(t=0,{\bm r})\sin\theta/2$ where the phase $\theta$ uniformly distributes in the range of $[0,2\pi]$ with the $\theta$ being the random phase of the nucleated U(1) vacuum bubbles. Both of bubble radius and the bubble wall width are all determined through bounce solutions conducted by Anybubble~\cite{Masoumi:2017trx} and FindBounce~\cite{Guada:2020xnz}.
 We consider that bubbles exponentially nucleate in the
symmetric phase, with the nucleation
probability being
\begin{equation}\label{eq:exp_nuc}
p(t)=p_f\exp[\beta(t-t_f)]\text,
\end{equation}
where $\beta = - \left. d \ln p(t) /d t \right|_{t_f} $ and $t_f$ is the
time at which the fraction of the universe in the symmetric phase is
$h(t_f)=1/e$~\cite{Enqvist:1991xw}.

%

\noindent{\it \bfseries  GWs production:}
During thermal first-order PT, there are mainly three GW contributions~\cite{Caprini:2019egz}: bubble collisions, sound waves, and turbulence. 
We calculate GWs by including all scalars contributions involved in PTs and  the evolution process of cosmic strings, including bubble collisions and scalars oscillation during the dynamical EWSB process\footnote{Here we mention that conventional adopted envelope approximation is still under debate, see Ref.~\cite{Cutting:2018tjt,Cutting:2020nla,Konstandin:2017sat,Ellis:2020nnr,Lewicki:2020jiv,Lewicki:2020azd}.}. 
 The equation of motion of tensor perturbations $h_{ij}$ reads~\cite{GarciaBellido:2007af}
\begin{equation}\label{heq}
\ddot{h}_{ij} - \nabla^2 h_{ij} = 16 \pi G T^{\mathrm{TT}}_{ij}\;.
\end{equation}
Here the superscript $\mathrm{TT}$ denotes the transverse-traceless projection, and we include both two scalar fields contributions in the energy-momentum tensor 
\begin{equation}
	\begin{split}
T_{\mu\nu}=\partial_\mu \phi_i \partial_\nu \phi_i&-g_{\mu\nu}\frac{1}{2}(\partial \phi_i )^2\;,\label{ttt}
\end{split}
\end{equation}
for type-a and type-b PTs, and 
\begin{equation}
	\begin{split}
T_{\mu\nu}=\partial_\mu \Phi^\dag \partial_\nu \Phi&-g_{\mu\nu}\frac{1}{2}\rm{Re}[(\partial_i  \Phi^\dag \partial^i \Phi )]\;,\label{ttt}
\end{split}
\end{equation}
for type-c PT. When the scale of the global U(1) theory is close to electroweak scale, we also include the contributions from the Higgs field. We evolve equation of motions in Eq.~\ref{scaeq} and tensor perturbations in Eq.~\ref{heq} with a code based on {\it pystella}~\cite{Adshead:2019lbr}\footnote{github.com/zachjweiner/pystella}.
The energy spectrum of GWs is the GW energy density fraction per
logarithmic frequency interval,
\begin{equation}
	 \frac{d\Omega_{\mathrm{GW}}}{d\ln k}=\dfrac{1}{\rho_{c}}\frac{d\rho_\text{GW}(k)}{d\ln k}\;.
\end{equation}

\noindent{\it \bfseries Numerical results:\label{sec:Numerical-Simulation}}
Our simulations are performed on a cubic lattice with the resolution $L^3=256^{3} \Delta x$. The time spacing is chosen to be $\Delta t=\Delta x/5$. This choice of lattice spacing gives us enough resolution to ensure that we capture all the dynamics for GWs and cosmic strings. 
The mean bubble separation is obtained as $R_\star=(L^3/N_b)^{1/3}$ with $N_b$ being the number of the generated bubbles during the PT processes, which determines the Lorentz factor for bubbles to be $\gamma_\star=R_\star/(2 R_0)$, and the wall width as $L_w^\star=L_w/\gamma_\star$ when bubbles collide. Theoretically, the bubble nucleation rate is directly connected with the mean bubble separation, see Refs.~\cite{Hindmarsh:2019phv,Cutting:2020nla,Cutting:2018tjt,Hindmarsh:2013xza}. 
In this study, initial bubble radius and initial bubble wall thickness for the three types PTs under study are: $R_0(L_w)=15(3)  \Delta x$ for type-a PT, and $R_0(L_w)=12(5) \Delta x$ for type-b and type-c PTs. The lattice resolution are: $\Delta x=1.2/T_n$ for type-a PT, and $\Delta x=0.35/T_n$ for type-b and type-c PTs with $T_n$ being the bubble nucleation temperature. 

 \begin{figure}[!htp]
\begin{center}
\includegraphics[width=0.3\textwidth]{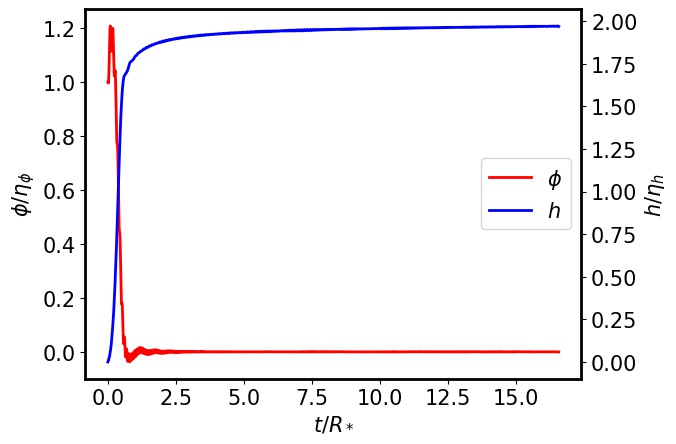}
\includegraphics[width=0.3\textwidth]{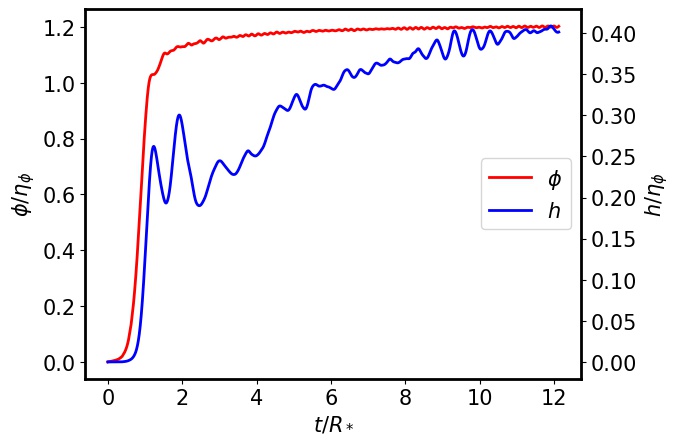}
\includegraphics[width=0.3\textwidth]{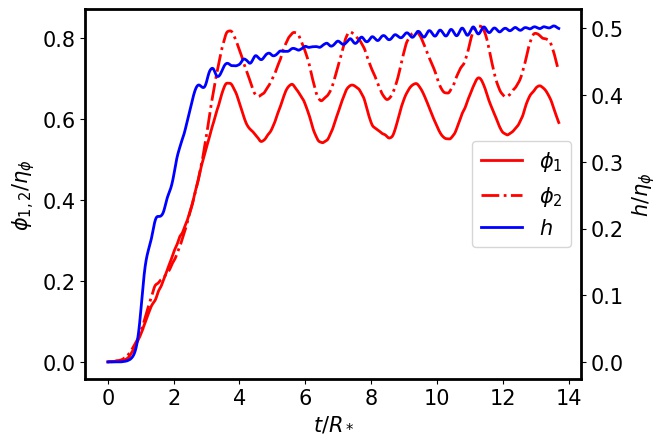}
\caption{Mean field evolutions during phase transition process of type-a (top), type-b (middle) and type-c (bottom) PTs.  } \label{fieldenv}
\end{center}
\end{figure} 

We first study how the EWSB occurs through the three types two-step PTs. For type-a PT,  we simulate the PT by adopting $\lambda_{h\phi}(\lambda_\phi)=0.66(1)$. To simulate the type-b PT and type-c PT as illustrated in the right plot of Fig.~\ref{ptpatterns}, we take model parameters as follows: $a= 0.125, b=0.0189977, \lambda_p=0.05$. In both type-a and type-b PTs, we consider the scale of $\phi$ to be close to the Higgs, where the interactions of the beyond SM sectors that trigger the EWSB are of order $\lambda_{h\phi}\sim\mathcal{O}(10^{-1})$ and $\lambda_{p}\sim\mathcal{O}(10^{-2})$ respectively. For the type-c PT, we take model parameters as type-b PT. Here, we note that a much smaller $\lambda_p$ can help to achieve the EWSB after breakdown of a high scale global U(1) theory.  In these cases, correspondingly,  the PT strength parameters (the released latent heats normalized by radiation energy) are: $\alpha= 0.013$ for type-a PT, and $\alpha= 0.081$ for type-b and type-c PTs.

In the top plot of Fig.~\ref{fieldenv}, we present the mean field values evolution of $\phi$ and $h$ to show the type-a PT process. 
The decrease of $\phi$ and the accompanying increase of $h$ indicate that the EWSB occurs around $t/R_\star\sim 1$ with the vacuum transiting as $(0,\langle \phi \rangle)\to (\langle h \rangle,0)$ (see the left plot of Fig.~\ref{ptpatterns}). The EWSB process occurs through electroweak bubbles expansion and merging with each other when the dark vacuum continues to shrink,  see Fig.~\ref{bubble_typea} of supplemental material for details on the dynamics of both
 dark vacuum bubbles and electroweak vacuum bubbles. Since evolution of both dark vacuum and electroweak vacuum can be described by bubble dynamics, we expect both of the two scalars of $\phi$ and $h$ can contribute to GWs production. 
 
In the middle plot of Fig.~\ref{fieldenv}, we present the mean field values evolution of $\phi$ and $h$ to describe the EWSB process in the type-b PT.
There, we find that the $\phi$ quickly gets VEV around $t/R_\star\sim 1$ when the CCT's vacuum bubbles expand and merge with each other. Then, the vacuum transits from the symmetric phase to the CCT's vacuum since the first-step PT is of first-order. Meanwhile, the second-step PT starts as the $h$ gradually increases during the first step PT, where the space distributions of the Higgs field values inherit the same shape as CCT's vacuum bubbles (see top panels of Fig.~\ref{bubble_typeb} in the supplemental material).  After the first-order PT in CCT, and the second step PT process continues and completes around $t/R_\star\sim 12$, which is pretty long in comparison with first-order PT. In this way, the EWSB occurs through dimensional transmutation process. 
During the whole PT process, the dynamic of the inherited bubbles is totally different from
the CCT's vacuum bubbles, for more details see Fig.~\ref{bubble_typeb} of the supplemental material. For the GW production, we include energy-momentum tensor contributions from both dynamics of $h$ and $\phi$. 

We show the first-step first-order PT process in type-c PT through the bottom plot of Fig.~\ref{fieldenv}. Here, we consider the U(1) scale is one order higher than the electroweak scale, such that the EWSB can occur with a Higgs-portal coupling $\lambda_p=0.05$ considering a global U(1) theory. The evolutions of the two components of the complex scalar $\Phi$ ($\phi_1$ and $\phi_2$) indeed reveal the mean phase evolution during CCT's vacuum bubbles nucleation, expansion, and percolation process. In this scenario, we found $\phi_1$ and $\phi_2$ stablize around $t/R_\star\sim 4$, which take a much longer time in comparison with the first-order PT processes of type-a and type-b PTs. During the PT process,  EWSB occurs when the Higgs field gets VEV through dimensional transmutation as type-b. We found that the cosmic strings emerge after the bubbles of $\Phi$ collide and merge with each other around $t/R_\star\sim1$ and disappear around $t/R_\star\sim 6$, see Fig.~\ref{strsep} of the supplemental materials for properties of cosmic strings during the type-c PT process. In Fig.~\ref{bubble_typec} of supplemental material, we present the phase distribution together with bubbles dynamics on 2d planes, where one can find vortex and anti-vortex pairs appear. The dynamics of cosmic strings can change the GW spectra as will be shown later. Our study further shows that, EWSB can be induced by a first-order PT of a  GUT scale U(1) theory with $\lambda_p=10^{-32}$.

 \begin{figure}[!htp]
\begin{center}
\includegraphics[width=0.35\textwidth]{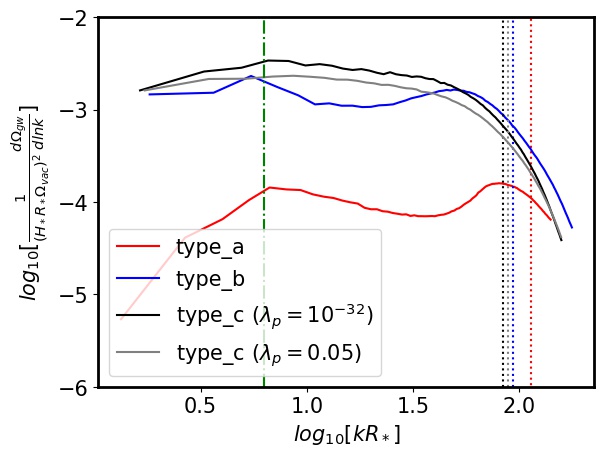}
\caption{GWs for three PT types. Bubble nucleation parameters are $\beta=0.023$ and $\beta=0.0098$ for type-a and type-b/type-c PTs with $p_f=\beta^4$. The solid-dashed and dotted lines mark the place of $R_\star$ and $L_w$. We apply cuttoffs in the UV frequencies to remove  numerical artifacts.  } \label{GWenv}
\end{center}
\end{figure}

We now turn to study the property of the GWs spectra generated from the PT processes of the three types, our results are shown in Fig.~\ref{GWenv}.  
Here, $\Omega_{vac}=\Delta\rho/\rho_c$ with $\Delta\rho$ being the latent heat released by the PT and $H_\star$ is the Hubble parameter at the time of $t/R_\star=1$.
The results show that GW spectra for both type-a and type-b PTs are all of broken power-law double-peak shapes, with the two peak frequency locations roughly corresponding to $ R_\star$ and $L_w$. 
The magnitude of the second peak located at  around the $L_w$ is almost the same as that of the $R_\star$. 
Different from the type-a PT, in type-b and type-c PTs,
there are less bubbles and bubble velocities are slighter larger. Therefore, the magnitude of GW spectra in type-b and type-c PTs are slightly higher than that of type-a PT. The bubble wall velocities are $v_b=0.95$ (and $0.83$) for type-b and type-c (and type-a PTs) respectively. 
For the two-step PT of type-a, our simulations show that bubbles dynamic of both dark vacuum and electroweak vacuum can contribute to GWs, and the contribution from bubble dynamics of $\phi$ is sub-dominate, see top plot of Fig.~\ref{expfiegw} in supplemental materials. 
For the type-b PT, the GW contributions from CCT's vacuum bubble dynamic is dominated, and the contributions of the dynamics of Higgs is sub-leading and mostly contribute to the peak of the GWs spectra corresponds to $L_w$, see middle panel of Fig.~\ref{expfiegw} in supplemental materials.  
For the case where the EWSB is induced by a first-order PT yielding spontaneous breaking of a global U(1) symmetry, the GW spectra of type-c PT is of plateau shape 
rather than double-peak shape as found in type-a and type-b PTs. In the case of $\lambda_p=10^{-32}$ (0.05), there are slightly more (less) CCT's vacuum bubbles nucleated and the bubble velocity is slightly larger (smaller), therefore the amplitude  of the GW spectra is lower (higher).
And for the type-c PT, we have cosmic strings formation and decay during the PT process, see Fig.~\ref{strsep} of supplemental material for cosmic strings evolution. Before the cosmic string decay to massive and goldstone particles, the cosmic string energy density (GW energy density) decreases (increases) as the 
proceeding of PT after bubbles collide with each other. The dynamics of cosmic strings amplifies  the strength of GWs a lot after the mean bubble separation time ($t>R_\star$), as can be found in Fig.~\ref{expfiegw} of supplemental material.

\noindent{\it \bfseries Conclusions and discussions:}
\label{sec:conclusions}
In this Letter, we performed 3d numerical simulation of the EWSB induced by the beyond SM sector through two-step PTs for the first time, where these models are well motivated by the dark matter, baryogenesis, gravitational waves, strong CP problem, and gauge hierarchy problem etc. 
When the beyond SM sectors are around the weak scale and interact weakly with the SM through the Higgs-portal, 
our simulation yields  two-peak GW spectra with broken power-law,  with the locations of two peak frequencies corresponding to the mean bubble separation and the bubble wall width, respectively.
For a high scale first-order PT of a global U(1) theory proceeded by EWSB through dimensional transmutation, we found that the  resulting GW spectra are of plateau shapes associated with the production and decay of cosmic strings rather than a broken power-shape, and the cutoff on the high peak frequency comes from the bubble wall width. 
Our observations show that GW observations can help to differentiate different classes of beyond SM physics. In the early Universe, when the underlying theory inducing  the EWSB obeys (does not obey) a global U(1) symmetry, the produced GW spectrum  is of  a plateau shape (broken double-peak power-law shape).

\noindent{\it \bfseries Acknowledgements}
We thank John T. Giblin, Marek Lewicki, Daniel Cutting, David Weir, Adrien Florio, Zach Weiner, Daniel G. Figueroa, Michael J. Ramsey-Musolf, Xuefeng Zhang, Shaojiang Wang  and Huaike Guo for communications. 
The work of Ligong Bian is supported by the National Key Research and Development Program of China Grant No. 2021YFC2203004, the National Natural Science Foundation of China under the grants Nos.12075041, 12047564, and the Fundamental Research Funds for the Central Universities of China (No. 2021CDJQY-011 and No. 2020CDJQY-Z003), and Chongqing Natural Science Foundation (Grants No.cstc2020jcyj-msxmX0814). RGC is supported in part by the National Key Research and Development Program of China Grant Nos. 2020YFC22015092 and 2021YFA0718304 and the National Natural Science Foundation of China under the grant Nos.11821505 and 11991052. 
\begin{figure}[!htp]
\begin{center}
\includegraphics[width=0.2\textwidth]{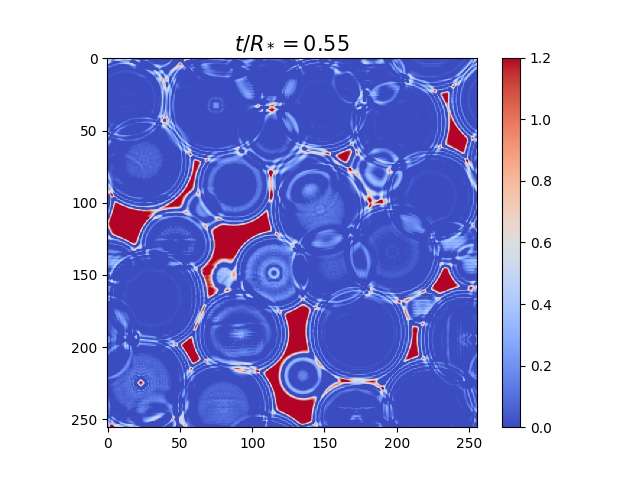}
\includegraphics[width=0.2\textwidth]{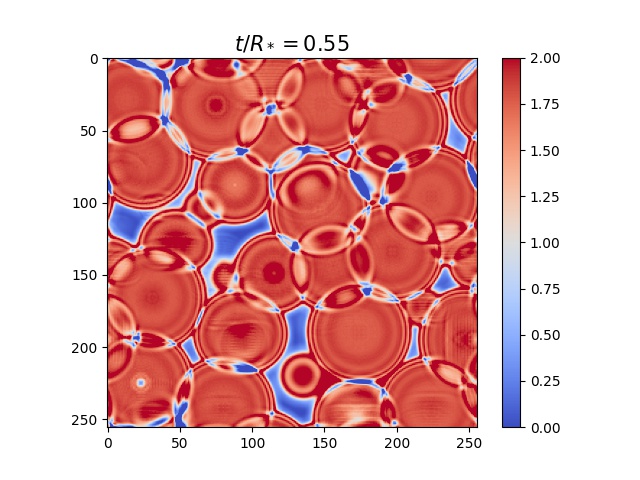}
\includegraphics[width=0.2\textwidth]{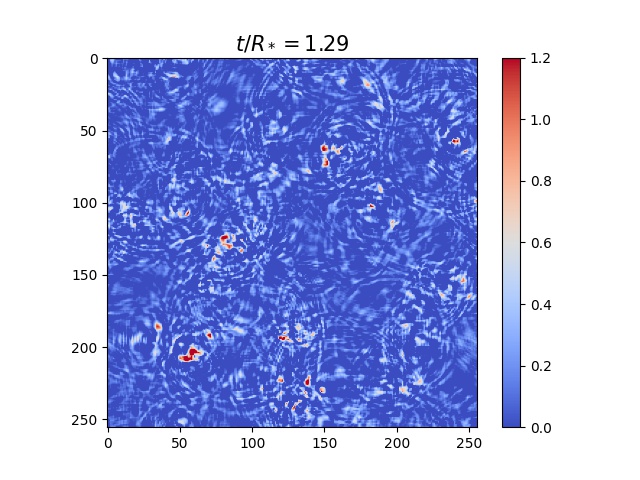}
\includegraphics[width=0.2\textwidth]{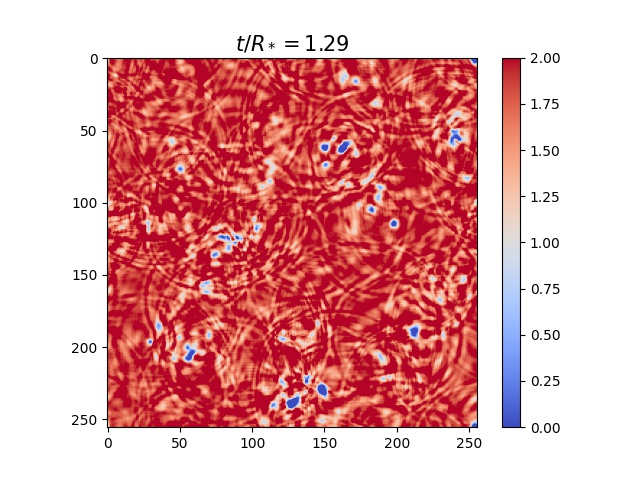}
\includegraphics[width=0.2\textwidth]{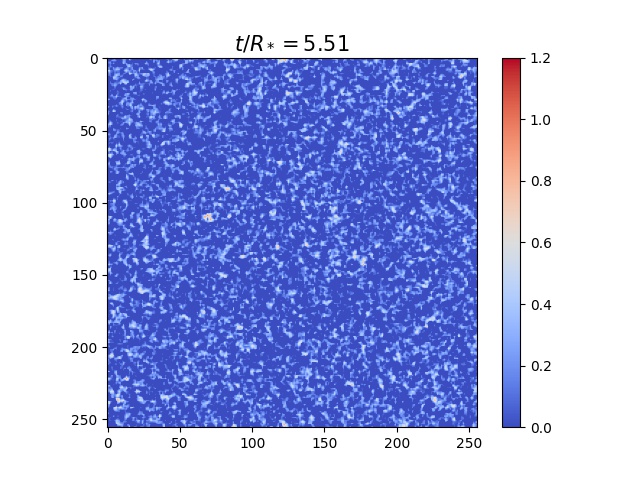}
\includegraphics[width=0.2\textwidth]{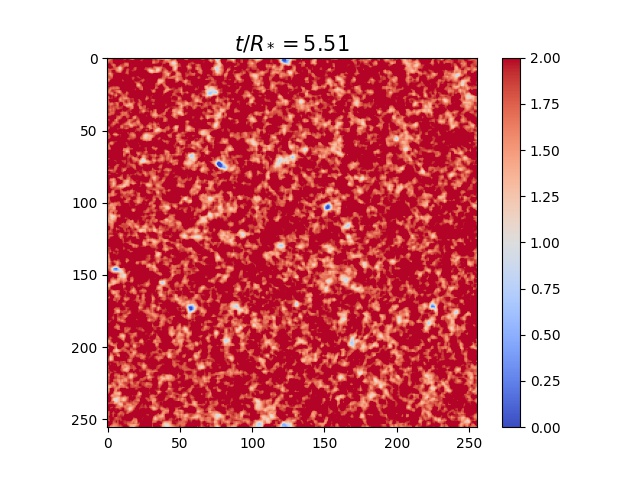}
\caption{For illustration, we plot the 2d slices for bubble dynamics of $\phi$ (left panels) and $h$ (right panels) at different time during the second-step first-order PT for the type-a PT. The color bar represents the normalized magnitude of the $h,\phi$ fields at each lattice points.} \label{bubble_typea}
\end{center}
\end{figure}

\noindent{\it \bfseries  Supplemental material}
In Fig.~\ref{bubble_typea}, we present three time slices during the expansion and collision
of the nucleated bubbles for the  type-a phase transition. The left panels represent the bubble dynamics when the first-order PT of 
type-a occurs (see Fig.~\ref{ptpatterns}). As depicted in the top two plots, we firstly have the electroweak vacuum  inside the bubbles which is the false vacuum of $\phi$ (dark vacuum), and region outside the vacuum bubbles is the dark vacuum rather than the electroweak vacuum. 
Around $t/R_\star\sim 1$, bubbles start to collide and merge with each other, where Higgs gets VEV through the electroweak vacuum eaten the dark vacuum, therefore we found that in most places the normalized value of $h$ (and $\phi$) approaches to 1 (and 0), see middle plots. With the proceeding of the PT, we further present a later time slice at $t>R_\star$ to show that the transition from the dark vacuum to the electroweak vacuum will complete and the EWSB occurs.  

\begin{figure}[!htp]
\begin{center}
\includegraphics[width=0.2\textwidth]{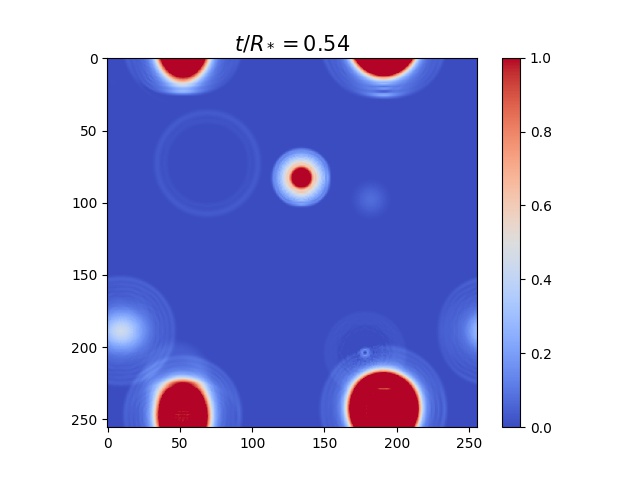}
\includegraphics[width=0.2\textwidth]{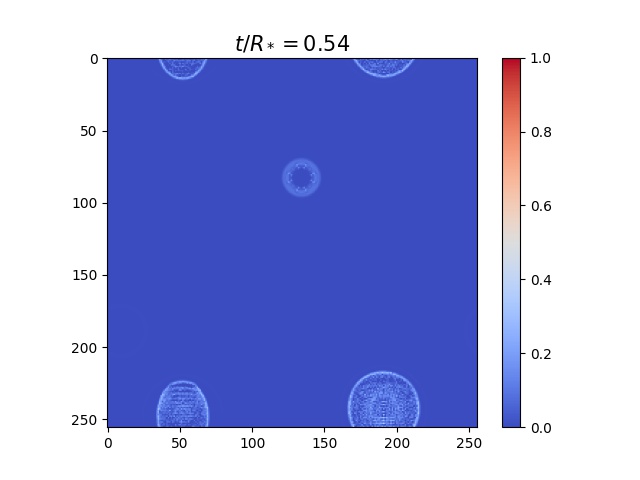}
\includegraphics[width=0.2\textwidth]{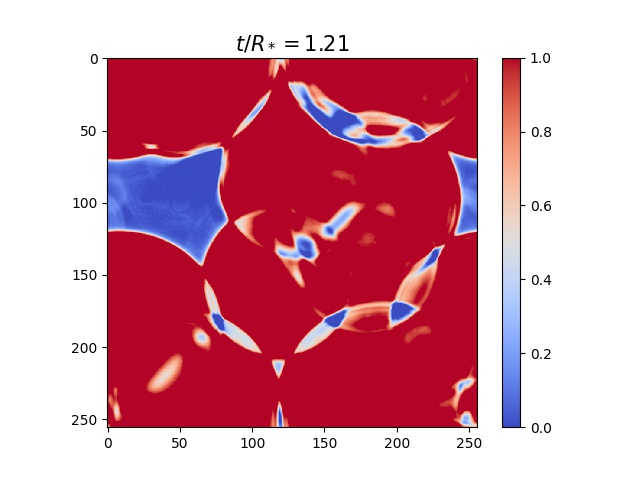}
\includegraphics[width=0.2\textwidth]{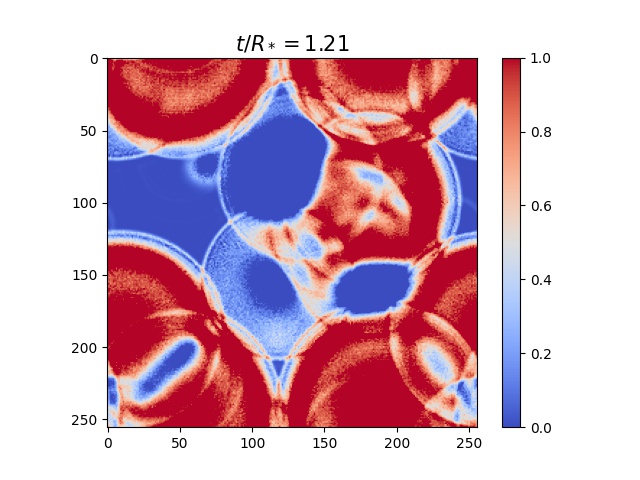}
\includegraphics[width=0.2\textwidth]{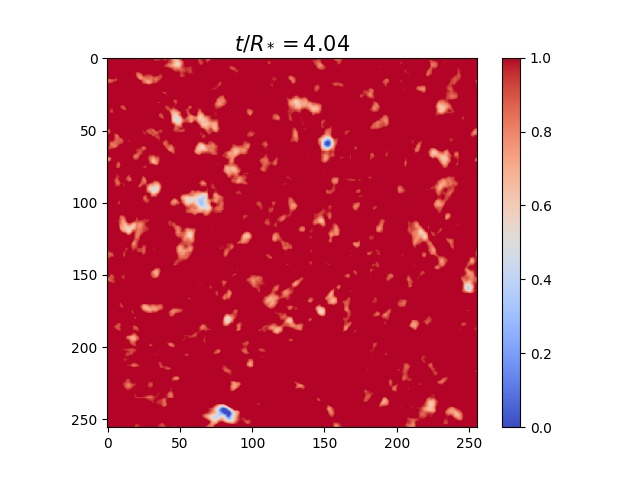}
\includegraphics[width=0.2\textwidth]{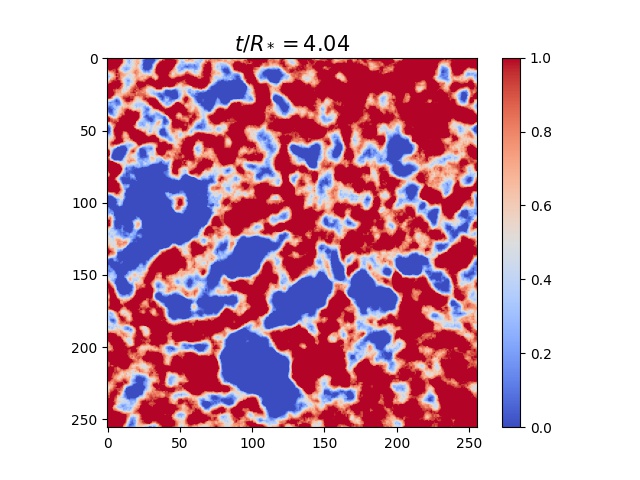}
\caption{The 2d slices for bubbles dynamic (left panels) and Higgs field values (right panels) at different time during the type-b PT of simultaneous bubble nucleation case. Here, the color bar is the same as Fig.~\ref{bubble_typea}. } \label{bubble_typeb}
\end{center}
\end{figure}

Fig.~\ref{bubble_typeb} shows the bubble dynamics for the type-b PT. Top two plots show that when the bubbles of CCT's vacuum nucleate, the distribution of Higgs field value also shares the same bubble shape at the early stage since the Higgs gets VEV through the dimensional transmutation after $\phi$ gets VEV. In the second-row plots, we find that, at the time of $t/R_\star=1.21$, the distribution of Higgs field value is different from that of $\phi$ for  the same reason, one needs to keep in mind that it's the natural result of dimensional transmutation. In the bottom plots, one can find that after the vacuum falls into the CCT's vacuum (i.e., after the merge of  CCT's vacuum bubbles), the EWSB finishes later as can also be found in mean field value evolution shown in the middle panel of Fig.~\ref{fieldenv}.

\begin{figure}[!htp]
\begin{center}
\includegraphics[width=0.2\textwidth]{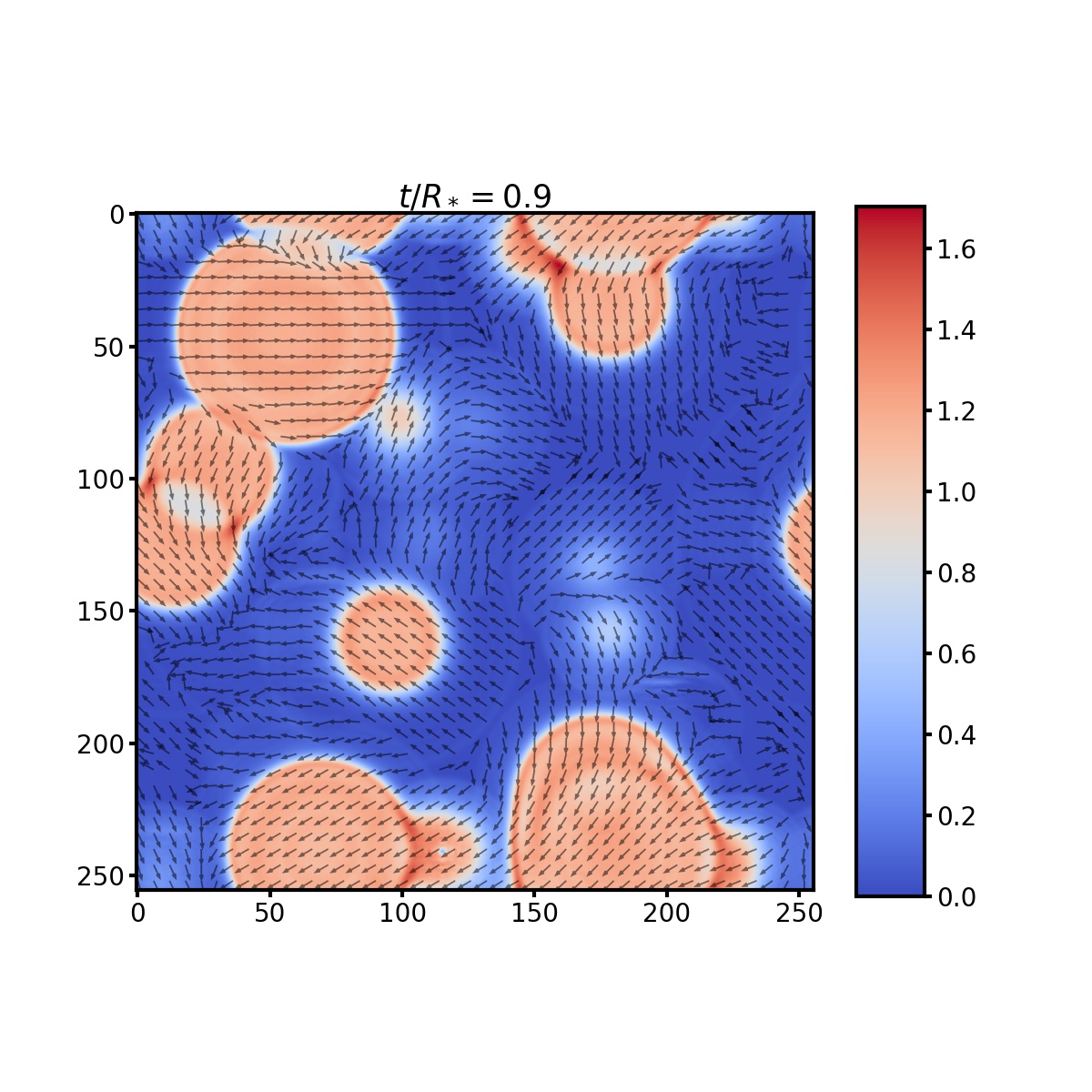}
\includegraphics[width=0.2\textwidth]{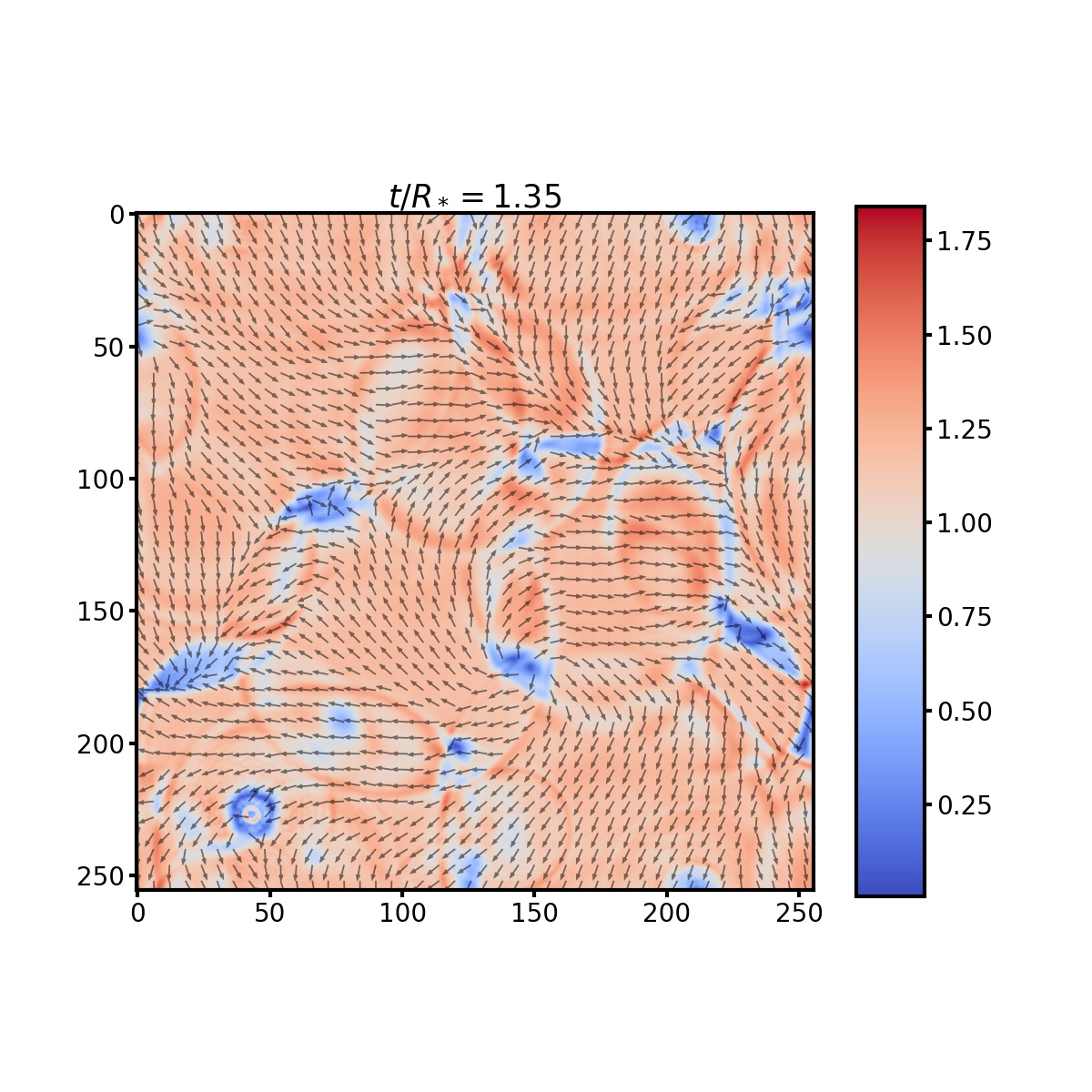}
\includegraphics[width=0.2\textwidth]{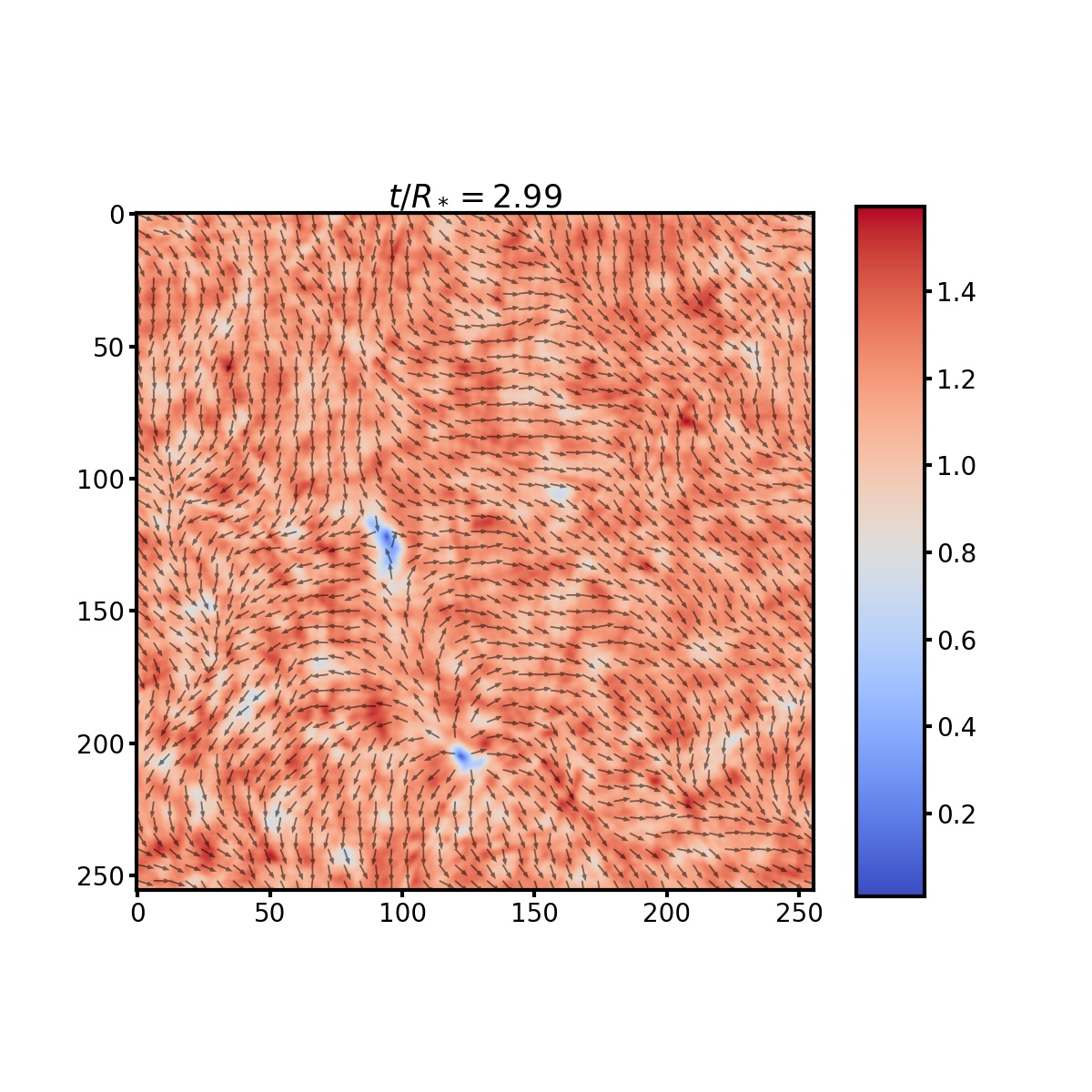}
\includegraphics[width=0.2\textwidth]{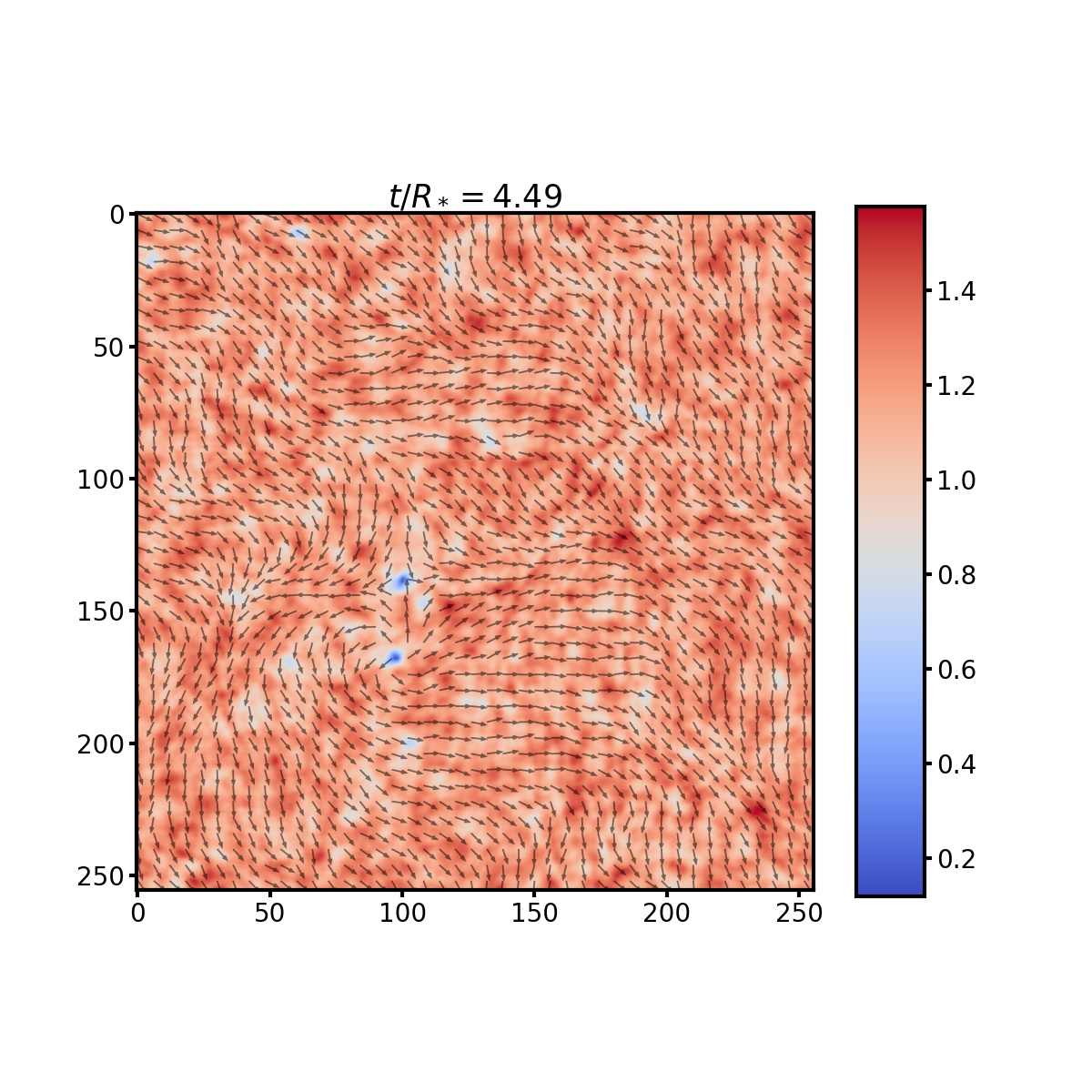}
\caption{The bubble dynamics and phase distribution at different time in the type-c PT. The color bar here is to represent the magnitude of the complex field at each lattice point, and the arrows represent phases at each place.} \label{bubble_typec}
\end{center}
\end{figure}

For the first-step of type-c PT, during the first-order PT process, we find that cosmic strings form after the U(1) vacuum bubbles collide with each other at $t/R_\star >1$. We present the 2d slices of bubble dynamics and phase variations in Fig.~\ref{bubble_typec}, where we can find that vortex and anti-vortex pairs are formed after bubbles collide with each other (see bottom two plots), where the false vacua are trapped inside the true vacua. As shown in the left panel, before bubbles collide with each other, the phase in each bubble is the same. After bubbles start to collide with each other, the phases start to redistribute as can be found in the top-right plot. Some time later after $t/R_\star>1$, vortex and anti-vortex pairs appear as can be found in the phase diagram shown in the bottom two plots.

 \begin{figure}[!htp]
\begin{center}
\includegraphics[width=0.2\textwidth]{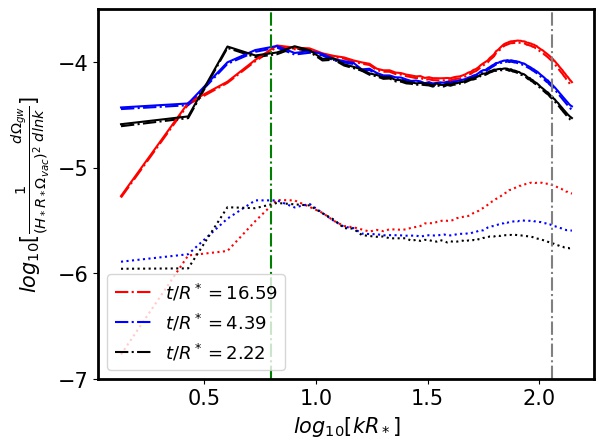}
\includegraphics[width=0.2\textwidth]{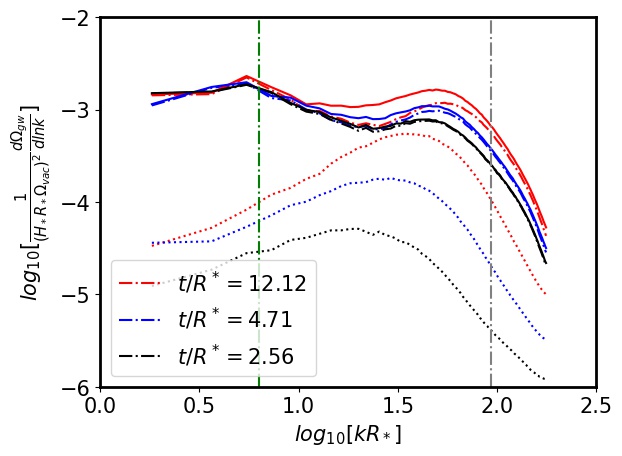}
\includegraphics[width=0.2\textwidth]{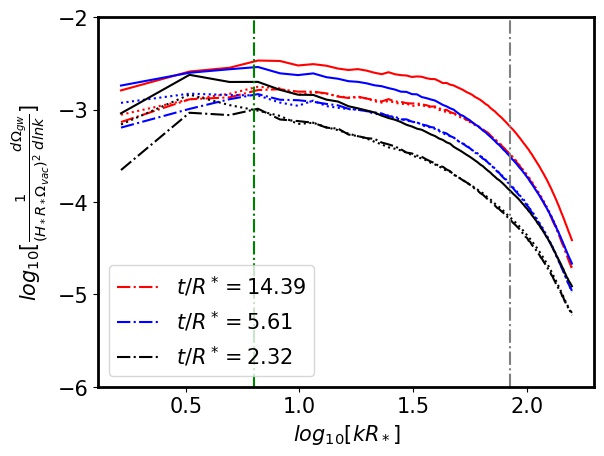}
\includegraphics[width=0.2\textwidth]{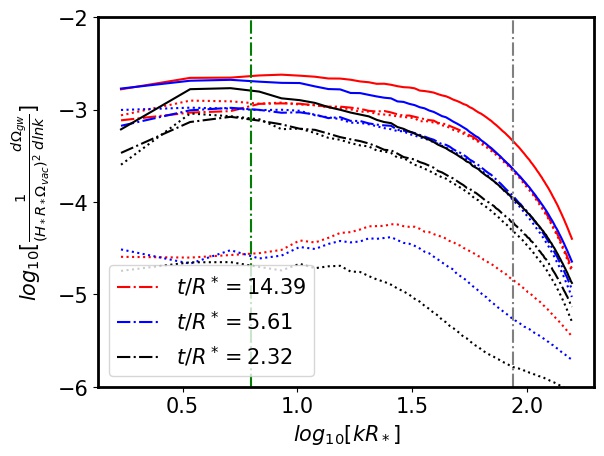}
\caption{GWs spectra  for three types of two-step PTs, type-a (top-left), type-b (top-right), type-c PT with $\lambda_p=10^{-32}$ (bottom-left), type-c PT with $\lambda_p=0.05$ (bottom-right). Bubble nucleation parameters are taken as: $\beta=0.023$ and $p_f=\beta^4$ for type-a PT, and $\beta=0.0098$ and $p_f=\beta^4$ for type-b and type-c PTs. The dash-dot and dotted curves for type-a ( type-b) represent contributions of $h$ and $\phi$ ($\phi$ and h). For the type-c, the GWs from energy-momentum tensor of real and imaginary fields are depicted by dash-dot and dotted curves, and the dotted curves indicate GW contributions from dynamics of the Higgs. The total contributions from all scalars are plotted with solid curves.
The green and grey dashed vertical dash-dot lines are used to mark the place of $R_\star$ and $L_w$ on the GWs spectra.} \label{expfiegw}
\end{center}
\end{figure}

The top-left and top-right plots of Fig.~\ref{expfiegw} show that the GWs spectra for the type-a and type-b PT types are all of broken power-law double-peak shapes. 
In the type-a PT case, the GWs contributions from the bubble dynamics of dark vacuum is one-order in magnitude  smaller than that of the electroweak vacuum. The contribution from the CCT's bubble dynamics dominates the GWs production in the type-b PT, where the dynamics of Higgs mostly affects the GWs spectra around the second peak locating at around the scale of the bubble wall width. 
Here, we note that the amplitude  of GWs of type-b PT is slightly higher than that of type-a PT due to the bubble wall velocity of type-b PT ($v_b=0.95$) is slightly larger than that of type-a PT ($v_b=0.83$). 
The bottom plot demonstrates that, for the case of type-c PTs, the GW spectra with cosmic strings formation
is of a plateau shape rather than  the double peak broken power-law shape. The GWs contribution from the dynamics of the Higgs during the PT is negligible for both large and small Higgs-portal coupling $\lambda_p$. 

Follow the convention of~\cite{Saurabh:2020pqe},  the energy density stored in the cosmic string is given by 
\be
{\cal E} = \frac{1}{2} |\partial_t \Phi |^2+\frac{1}{2} |\nabla \Phi |^2+ V_{cct}(\Phi,T)\;,
\ee
Here, $\Phi\equiv \rho \, \exp(i\theta )$ with 
$\rho=\sqrt{\phi_1^2+\phi_2^2}$ and $\theta=\arcsin[\frac{\phi_1}{\sqrt{\phi_1^2+\phi_2^2}}]$.
We then obtain
\be
{\cal E} \equiv {\cal E}_\rho + {\cal E}_\theta ,
\ee
where the energy density of massive modes ($\rho$) is 
\be
{\cal E}_\rho = 
\frac{1}{2} (\partial_t \rho )^2+\frac{1}{2} (\nabla \rho )^2 
+ V_{ cct}(\rho) ,
\label{Erho}
\ee
and that of the Goldstone modes ($\theta$) is defined as
\be
{\cal E}_\theta = 
\frac{\rho^2}{2} \left [ (\partial_t \theta )^2+ (\nabla \theta)^2 \right ] .
\label{Ealpha}
\ee
Here, we note that this $\theta$ is not the phase associated with nucleated bubbles.

 \begin{figure}[!htp]
\begin{center}
\includegraphics[width=0.23\textwidth]{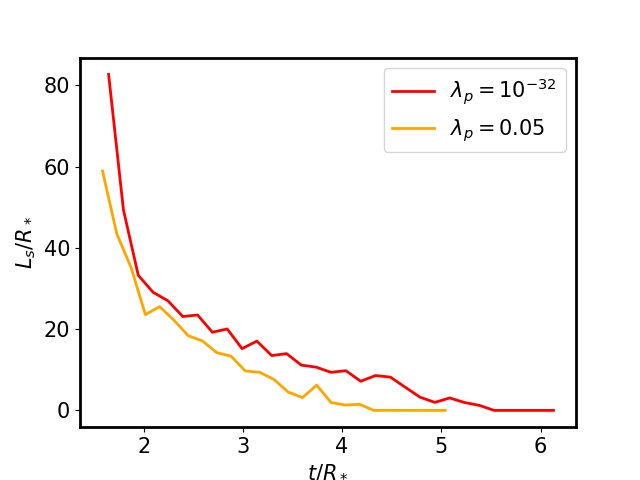}
\includegraphics[width=0.20\textwidth]{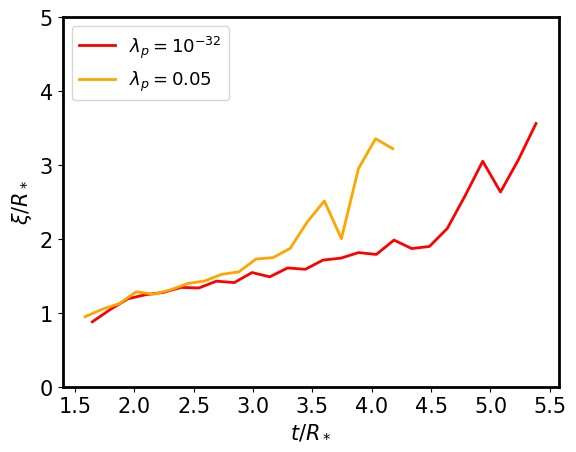}
\includegraphics[width=0.21\textwidth]{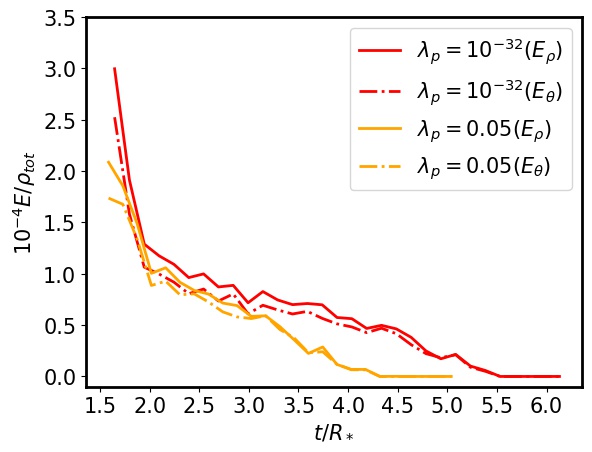}
\includegraphics[width=0.22\textwidth]{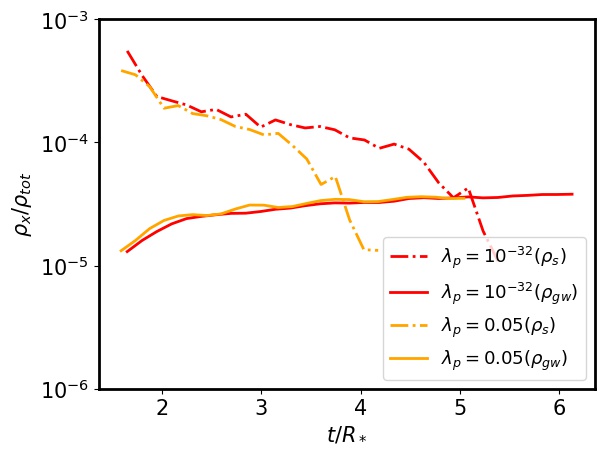}
\caption{Properties of cosmic strings during the type-c PT. Top-left: String length evolutions; Top-right: The corresponding evolutions of mean string separation;
Bottom-left: The string energy density of $E_\rho$(solid lines) and $E_\theta$ (dash-dot lines); Bottom-right: The GW (solid lines) and string energy density (dash-dot lines) evolutions during the PTs.} \label{strsep}
\end{center}
\end{figure}

As shown in Fig.~\ref{strsep}, our simulations indicate that cosmic strings  start to form after vacuum bubbles collide with each other, i.e., $t/R_\star>1$, and disappear depending  on the number of bubble and bubble velocities. Here, we take $|\Phi|/\eta_\Phi<0.1$ to identify cosmic strings.
Fig.~\ref{strsep} demonstrates that the cosmic string length ($L_s$) decreases (top-left panel), and the mean string separation (top-right panel) exponentially increases until cosmic strings disappear. The mean string separation here is defined as $V/L_s$ and its minimum is around $R_\star$ when cosmic strings are formed at the beginning. The string length and the mean string separation for $\lambda_p=0.05$ cases are smaller than  those in the case of $\lambda_{p}=10^{-32}$ due to there are much less bubbles nucleated during the PT in the former case, we have $N_b=50$ and $N_b=56$ for the cases of $\lambda_p=0.05$ and $\lambda_p=10^{-32}$.
To illustrate the relation between the bubble dynamics and the radiation mode of cosmic strings, we present the comic strings energy densities (including massive mode and goldstone mode) in the bottom-left panel of Fig.~\ref{strsep}.  Here, we find that the energy density of the massive modes is slightly larger than that of the goldstone modes. The bottom-right panel depicts that cosmic strings can lose energy through particle radiations and gravitational wave radiations. The cosmic strings energy density is much larger than the GW energy density when cosmic strings are formed, and the cosmic string energy density (and GW energy density) decreases (increases) before cosmic strings disappear for the case of large $\gamma_\star$ with ultra-relativistic bubble velocity under study.  For the study on particle radiations after a second order phase transition of some U(1) theories, we refer to Refs.~\cite{Vilenkin:1982ks,Saurabh:2020pqe}.

\bibliographystyle{apsrev}

\bibliography{GWBPT}

\begin{thebibliography}{99}
\expandafter\ifx\csname natexlab\endcsname\relax\def\natexlab#1{#1}\fi
\expandafter\ifx\csname bibnamefont\endcsname\relax
  \def\bibnamefont#1{#1}\fi
\expandafter\ifx\csname bibfnamefont\endcsname\relax
  \def\bibfnamefont#1{#1}\fi
\expandafter\ifx\csname citenamefont\endcsname\relax
  \def\citenamefont#1{#1}\fi
\expandafter\ifx\csname url\endcsname\relax
  \def\url#1{\texttt{#1}}\fi
\expandafter\ifx\csname urlprefix\endcsname\relax\def\urlprefix{URL }\fi
\providecommand{\bibinfo}[2]{#2}
\providecommand{\eprint}[2][]{\url{#2}}

\bibitem[{\citenamefont{D'Onofrio et~al.}(2014)\citenamefont{D'Onofrio,
  Rummukainen, and Tranberg}}]{DOnofrio:2014rug}
\bibinfo{author}{\bibfnamefont{M.}~\bibnamefont{D'Onofrio}},
  \bibinfo{author}{\bibfnamefont{K.}~\bibnamefont{Rummukainen}},
  \bibnamefont{and} \bibinfo{author}{\bibfnamefont{A.}~\bibnamefont{Tranberg}},
  \bibinfo{journal}{Phys. Rev. Lett.} \textbf{\bibinfo{volume}{113}},
  \bibinfo{pages}{141602} (\bibinfo{year}{2014}), \eprint{1404.3565}.

\bibitem[{\citenamefont{Mazumdar and White}(2019)}]{Mazumdar:2018dfl}
\bibinfo{author}{\bibfnamefont{A.}~\bibnamefont{Mazumdar}} \bibnamefont{and}
  \bibinfo{author}{\bibfnamefont{G.}~\bibnamefont{White}},
  \bibinfo{journal}{Rept. Prog. Phys.} \textbf{\bibinfo{volume}{82}},
  \bibinfo{pages}{076901} (\bibinfo{year}{2019}), \eprint{1811.01948}.

\bibitem[{\citenamefont{Caldwell et~al.}(2022)}]{Caldwell:2022qsj}
\bibinfo{author}{\bibfnamefont{R.}~\bibnamefont{Caldwell}}
  \bibnamefont{et~al.}, in \emph{\bibinfo{booktitle}{{2022 Snowmass Summer
  Study}}} (\bibinfo{year}{2022}), \eprint{2203.07972}.

\bibitem[{\citenamefont{Abbott et~al.}(2016)}]{Abbott:2016blz}
\bibinfo{author}{\bibfnamefont{B.~P.} \bibnamefont{Abbott}}
  \bibnamefont{et~al.} (\bibinfo{collaboration}{LIGO Scientific, Virgo}),
  \bibinfo{journal}{Phys. Rev. Lett.} \textbf{\bibinfo{volume}{116}},
  \bibinfo{pages}{061102} (\bibinfo{year}{2016}), \eprint{1602.03837}.

\bibitem[{\citenamefont{Amaro-Seoane et~al.}(2017)}]{Audley:2017drz}
\bibinfo{author}{\bibfnamefont{P.}~\bibnamefont{Amaro-Seoane}}
  \bibnamefont{et~al.} (\bibinfo{collaboration}{LISA}) (\bibinfo{year}{2017}),
  \eprint{1702.00786}.

\bibitem[{\citenamefont{Ruan et~al.}(2020)\citenamefont{Ruan, Guo, Cai, and
  Zhang}}]{Guo:2018npi}
\bibinfo{author}{\bibfnamefont{W.-H.} \bibnamefont{Ruan}},
  \bibinfo{author}{\bibfnamefont{Z.-K.} \bibnamefont{Guo}},
  \bibinfo{author}{\bibfnamefont{R.-G.} \bibnamefont{Cai}}, \bibnamefont{and}
  \bibinfo{author}{\bibfnamefont{Y.-Z.} \bibnamefont{Zhang}},
  \bibinfo{journal}{Int. J. Mod. Phys. A} \textbf{\bibinfo{volume}{35}},
  \bibinfo{pages}{2050075} (\bibinfo{year}{2020}), \eprint{1807.09495}.

\bibitem[{\citenamefont{Luo et~al.}(2016)}]{Luo:2015ght}
\bibinfo{author}{\bibfnamefont{J.}~\bibnamefont{Luo}} \bibnamefont{et~al.}
  (\bibinfo{collaboration}{TianQin}), \bibinfo{journal}{Class. Quant. Grav.}
  \textbf{\bibinfo{volume}{33}}, \bibinfo{pages}{035010}
  (\bibinfo{year}{2016}), \eprint{1512.02076}.

\bibitem[{\citenamefont{Corbin and Cornish}(2006)}]{Corbin:2005ny}
\bibinfo{author}{\bibfnamefont{V.}~\bibnamefont{Corbin}} \bibnamefont{and}
  \bibinfo{author}{\bibfnamefont{N.~J.} \bibnamefont{Cornish}},
  \bibinfo{journal}{Class. Quant. Grav.} \textbf{\bibinfo{volume}{23}},
  \bibinfo{pages}{2435} (\bibinfo{year}{2006}), \eprint{gr-qc/0512039}.

\bibitem[{\citenamefont{Yagi and Seto}(2011)}]{Yagi:2011wg}
\bibinfo{author}{\bibfnamefont{K.}~\bibnamefont{Yagi}} \bibnamefont{and}
  \bibinfo{author}{\bibfnamefont{N.}~\bibnamefont{Seto}},
  \bibinfo{journal}{Phys. Rev. D} \textbf{\bibinfo{volume}{83}},
  \bibinfo{pages}{044011} (\bibinfo{year}{2011}), \bibinfo{note}{[Erratum:
  Phys.Rev.D 95, 109901 (2017)]}, \eprint{1101.3940}.

\bibitem[{\citenamefont{Xue et~al.}(2021)}]{Xue:2021gyq}
\bibinfo{author}{\bibfnamefont{X.}~\bibnamefont{Xue}} \bibnamefont{et~al.},
  \bibinfo{journal}{Phys. Rev. Lett.} \textbf{\bibinfo{volume}{127}},
  \bibinfo{pages}{251303} (\bibinfo{year}{2021}), \eprint{2110.03096}.

\bibitem[{\citenamefont{Arzoumanian et~al.}(2021)}]{NANOGrav:2021flc}
\bibinfo{author}{\bibfnamefont{Z.}~\bibnamefont{Arzoumanian}}
  \bibnamefont{et~al.} (\bibinfo{collaboration}{NANOGrav}),
  \bibinfo{journal}{Phys. Rev. Lett.} \textbf{\bibinfo{volume}{127}},
  \bibinfo{pages}{251302} (\bibinfo{year}{2021}), \eprint{2104.13930}.

\bibitem[{\citenamefont{Romero et~al.}(2021)\citenamefont{Romero, Martinovic,
  Callister, Guo, Mart\'\i{}nez, Sakellariadou, Yang, and
  Zhao}}]{Romero:2021kby}
\bibinfo{author}{\bibfnamefont{A.}~\bibnamefont{Romero}},
  \bibinfo{author}{\bibfnamefont{K.}~\bibnamefont{Martinovic}},
  \bibinfo{author}{\bibfnamefont{T.~A.} \bibnamefont{Callister}},
  \bibinfo{author}{\bibfnamefont{H.-K.} \bibnamefont{Guo}},
  \bibinfo{author}{\bibfnamefont{M.}~\bibnamefont{Mart\'\i{}nez}},
  \bibinfo{author}{\bibfnamefont{M.}~\bibnamefont{Sakellariadou}},
  \bibinfo{author}{\bibfnamefont{F.-W.} \bibnamefont{Yang}}, \bibnamefont{and}
  \bibinfo{author}{\bibfnamefont{Y.}~\bibnamefont{Zhao}},
  \bibinfo{journal}{Phys. Rev. Lett.} \textbf{\bibinfo{volume}{126}},
  \bibinfo{pages}{151301} (\bibinfo{year}{2021}), \eprint{2102.01714}.

\bibitem[{\citenamefont{Jiang and Huang}(2022)}]{Jiang:2022mzt}
\bibinfo{author}{\bibfnamefont{Y.}~\bibnamefont{Jiang}} \bibnamefont{and}
  \bibinfo{author}{\bibfnamefont{Q.-G.} \bibnamefont{Huang}}
  (\bibinfo{year}{2022}), \eprint{2203.11781}.

\bibitem[{\citenamefont{Caprini et~al.}(2016)}]{Caprini:2015zlo}
\bibinfo{author}{\bibfnamefont{C.}~\bibnamefont{Caprini}} \bibnamefont{et~al.},
  \bibinfo{journal}{JCAP} \textbf{\bibinfo{volume}{04}}, \bibinfo{pages}{001}
  (\bibinfo{year}{2016}), \eprint{1512.06239}.

\bibitem[{\citenamefont{Caprini et~al.}(2020)}]{Caprini:2019egz}
\bibinfo{author}{\bibfnamefont{C.}~\bibnamefont{Caprini}} \bibnamefont{et~al.},
  \bibinfo{journal}{JCAP} \textbf{\bibinfo{volume}{03}}, \bibinfo{pages}{024}
  (\bibinfo{year}{2020}), \eprint{1910.13125}.

\bibitem[{\citenamefont{Kuzmin et~al.}(1985)\citenamefont{Kuzmin, Rubakov, and
  Shaposhnikov}}]{Kuzmin:1985mm}
\bibinfo{author}{\bibfnamefont{V.~A.} \bibnamefont{Kuzmin}},
  \bibinfo{author}{\bibfnamefont{V.~A.} \bibnamefont{Rubakov}},
  \bibnamefont{and} \bibinfo{author}{\bibfnamefont{M.~E.}
  \bibnamefont{Shaposhnikov}}, \bibinfo{journal}{Phys. Lett. B}
  \textbf{\bibinfo{volume}{155}}, \bibinfo{pages}{36} (\bibinfo{year}{1985}).

\bibitem[{\citenamefont{Shaposhnikov}(1986)}]{Shaposhnikov:1986jp}
\bibinfo{author}{\bibfnamefont{M.~E.} \bibnamefont{Shaposhnikov}},
  \bibinfo{journal}{JETP Lett.} \textbf{\bibinfo{volume}{44}},
  \bibinfo{pages}{465} (\bibinfo{year}{1986}).

\bibitem[{\citenamefont{Shaposhnikov}(1987)}]{Shaposhnikov:1987tw}
\bibinfo{author}{\bibfnamefont{M.~E.} \bibnamefont{Shaposhnikov}},
  \bibinfo{journal}{Nucl. Phys. B} \textbf{\bibinfo{volume}{287}},
  \bibinfo{pages}{757} (\bibinfo{year}{1987}).

\bibitem[{\citenamefont{Morrissey and Ramsey-Musolf}(2012)}]{Morrissey:2012db}
\bibinfo{author}{\bibfnamefont{D.~E.} \bibnamefont{Morrissey}}
  \bibnamefont{and} \bibinfo{author}{\bibfnamefont{M.~J.}
  \bibnamefont{Ramsey-Musolf}}, \bibinfo{journal}{New J. Phys.}
  \textbf{\bibinfo{volume}{14}}, \bibinfo{pages}{125003}
  (\bibinfo{year}{2012}), \eprint{1206.2942}.

\bibitem[{\citenamefont{Patel and Ramsey-Musolf}(2013)}]{Patel:2012pi}
\bibinfo{author}{\bibfnamefont{H.~H.} \bibnamefont{Patel}} \bibnamefont{and}
  \bibinfo{author}{\bibfnamefont{M.~J.} \bibnamefont{Ramsey-Musolf}},
  \bibinfo{journal}{Phys. Rev. D} \textbf{\bibinfo{volume}{88}},
  \bibinfo{pages}{035013} (\bibinfo{year}{2013}), \eprint{1212.5652}.

\bibitem[{\citenamefont{Blinov et~al.}(2015)\citenamefont{Blinov, Kozaczuk,
  Morrissey, and Tamarit}}]{Blinov:2015sna}
\bibinfo{author}{\bibfnamefont{N.}~\bibnamefont{Blinov}},
  \bibinfo{author}{\bibfnamefont{J.}~\bibnamefont{Kozaczuk}},
  \bibinfo{author}{\bibfnamefont{D.~E.} \bibnamefont{Morrissey}},
  \bibnamefont{and} \bibinfo{author}{\bibfnamefont{C.}~\bibnamefont{Tamarit}},
  \bibinfo{journal}{Phys. Rev. D} \textbf{\bibinfo{volume}{92}},
  \bibinfo{pages}{035012} (\bibinfo{year}{2015}), \eprint{1504.05195}.

\bibitem[{\citenamefont{Inoue et~al.}(2016)\citenamefont{Inoue, Ovanesyan, and
  Ramsey-Musolf}}]{Inoue:2015pza}
\bibinfo{author}{\bibfnamefont{S.}~\bibnamefont{Inoue}},
  \bibinfo{author}{\bibfnamefont{G.}~\bibnamefont{Ovanesyan}},
  \bibnamefont{and} \bibinfo{author}{\bibfnamefont{M.~J.}
  \bibnamefont{Ramsey-Musolf}}, \bibinfo{journal}{Phys. Rev. D}
  \textbf{\bibinfo{volume}{93}}, \bibinfo{pages}{015013}
  (\bibinfo{year}{2016}), \eprint{1508.05404}.

\bibitem[{\citenamefont{Ramsey-Musolf et~al.}(2018)\citenamefont{Ramsey-Musolf,
  Winslow, and White}}]{Ramsey-Musolf:2017tgh}
\bibinfo{author}{\bibfnamefont{M.~J.} \bibnamefont{Ramsey-Musolf}},
  \bibinfo{author}{\bibfnamefont{P.}~\bibnamefont{Winslow}}, \bibnamefont{and}
  \bibinfo{author}{\bibfnamefont{G.}~\bibnamefont{White}},
  \bibinfo{journal}{Phys. Rev. D} \textbf{\bibinfo{volume}{97}},
  \bibinfo{pages}{123509} (\bibinfo{year}{2018}), \eprint{1708.07511}.

\bibitem[{\citenamefont{Xie et~al.}(2020)\citenamefont{Xie, Bian, and
  Wu}}]{Xie:2020bkl}
\bibinfo{author}{\bibfnamefont{K.-P.} \bibnamefont{Xie}},
  \bibinfo{author}{\bibfnamefont{L.}~\bibnamefont{Bian}}, \bibnamefont{and}
  \bibinfo{author}{\bibfnamefont{Y.}~\bibnamefont{Wu}}, \bibinfo{journal}{JHEP}
  \textbf{\bibinfo{volume}{12}}, \bibinfo{pages}{047} (\bibinfo{year}{2020}),
  \eprint{2005.13552}.

\bibitem[{\citenamefont{Jiang et~al.}(2016)\citenamefont{Jiang, Bian, Huang,
  and Shu}}]{Jiang:2015cwa}
\bibinfo{author}{\bibfnamefont{M.}~\bibnamefont{Jiang}},
  \bibinfo{author}{\bibfnamefont{L.}~\bibnamefont{Bian}},
  \bibinfo{author}{\bibfnamefont{W.}~\bibnamefont{Huang}}, \bibnamefont{and}
  \bibinfo{author}{\bibfnamefont{J.}~\bibnamefont{Shu}},
  \bibinfo{journal}{Phys. Rev. D} \textbf{\bibinfo{volume}{93}},
  \bibinfo{pages}{065032} (\bibinfo{year}{2016}), \eprint{1502.07574}.

\bibitem[{\citenamefont{Bian and Tang}(2018)}]{Bian:2018mkl}
\bibinfo{author}{\bibfnamefont{L.}~\bibnamefont{Bian}} \bibnamefont{and}
  \bibinfo{author}{\bibfnamefont{Y.-L.} \bibnamefont{Tang}},
  \bibinfo{journal}{JHEP} \textbf{\bibinfo{volume}{12}}, \bibinfo{pages}{006}
  (\bibinfo{year}{2018}), \eprint{1810.03172}.

\bibitem[{\citenamefont{Bian and Liu}(2019)}]{Bian:2018bxr}
\bibinfo{author}{\bibfnamefont{L.}~\bibnamefont{Bian}} \bibnamefont{and}
  \bibinfo{author}{\bibfnamefont{X.}~\bibnamefont{Liu}},
  \bibinfo{journal}{Phys. Rev. D} \textbf{\bibinfo{volume}{99}},
  \bibinfo{pages}{055003} (\bibinfo{year}{2019}), \eprint{1811.03279}.

\bibitem[{\citenamefont{Baker and Kopp}(2017)}]{Baker:2016xzo}
\bibinfo{author}{\bibfnamefont{M.~J.} \bibnamefont{Baker}} \bibnamefont{and}
  \bibinfo{author}{\bibfnamefont{J.}~\bibnamefont{Kopp}},
  \bibinfo{journal}{Phys. Rev. Lett.} \textbf{\bibinfo{volume}{119}},
  \bibinfo{pages}{061801} (\bibinfo{year}{2017}), \eprint{1608.07578}.

\bibitem[{\citenamefont{Baker et~al.}(2018)\citenamefont{Baker, Breitbach,
  Kopp, and Mittnacht}}]{Baker:2017zwx}
\bibinfo{author}{\bibfnamefont{M.~J.} \bibnamefont{Baker}},
  \bibinfo{author}{\bibfnamefont{M.}~\bibnamefont{Breitbach}},
  \bibinfo{author}{\bibfnamefont{J.}~\bibnamefont{Kopp}}, \bibnamefont{and}
  \bibinfo{author}{\bibfnamefont{L.}~\bibnamefont{Mittnacht}},
  \bibinfo{journal}{JHEP} \textbf{\bibinfo{volume}{03}}, \bibinfo{pages}{114}
  (\bibinfo{year}{2018}), \eprint{1712.03962}.

\bibitem[{\citenamefont{Chao et~al.}(2017)\citenamefont{Chao, Guo, and
  Shu}}]{Chao:2017vrq}
\bibinfo{author}{\bibfnamefont{W.}~\bibnamefont{Chao}},
  \bibinfo{author}{\bibfnamefont{H.-K.} \bibnamefont{Guo}}, \bibnamefont{and}
  \bibinfo{author}{\bibfnamefont{J.}~\bibnamefont{Shu}},
  \bibinfo{journal}{JCAP} \textbf{\bibinfo{volume}{09}}, \bibinfo{pages}{009}
  (\bibinfo{year}{2017}), \eprint{1702.02698}.

\bibitem[{\citenamefont{Schwaller}(2015)}]{Schwaller:2015tja}
\bibinfo{author}{\bibfnamefont{P.}~\bibnamefont{Schwaller}},
  \bibinfo{journal}{Phys. Rev. Lett.} \textbf{\bibinfo{volume}{115}},
  \bibinfo{pages}{181101} (\bibinfo{year}{2015}), \eprint{1504.07263}.

\bibitem[{\citenamefont{Jaeckel et~al.}(2016)\citenamefont{Jaeckel, Khoze, and
  Spannowsky}}]{Jaeckel:2016jlh}
\bibinfo{author}{\bibfnamefont{J.}~\bibnamefont{Jaeckel}},
  \bibinfo{author}{\bibfnamefont{V.~V.} \bibnamefont{Khoze}}, \bibnamefont{and}
  \bibinfo{author}{\bibfnamefont{M.}~\bibnamefont{Spannowsky}},
  \bibinfo{journal}{Phys. Rev. D} \textbf{\bibinfo{volume}{94}},
  \bibinfo{pages}{103519} (\bibinfo{year}{2016}), \eprint{1602.03901}.

\bibitem[{\citenamefont{Croon et~al.}(2018)\citenamefont{Croon, Sanz, and
  White}}]{Croon:2018erz}
\bibinfo{author}{\bibfnamefont{D.}~\bibnamefont{Croon}},
  \bibinfo{author}{\bibfnamefont{V.}~\bibnamefont{Sanz}}, \bibnamefont{and}
  \bibinfo{author}{\bibfnamefont{G.}~\bibnamefont{White}},
  \bibinfo{journal}{JHEP} \textbf{\bibinfo{volume}{08}}, \bibinfo{pages}{203}
  (\bibinfo{year}{2018}), \eprint{1806.02332}.

\bibitem[{\citenamefont{Breitbach et~al.}(2019)\citenamefont{Breitbach, Kopp,
  Madge, Opferkuch, and Schwaller}}]{Breitbach:2018ddu}
\bibinfo{author}{\bibfnamefont{M.}~\bibnamefont{Breitbach}},
  \bibinfo{author}{\bibfnamefont{J.}~\bibnamefont{Kopp}},
  \bibinfo{author}{\bibfnamefont{E.}~\bibnamefont{Madge}},
  \bibinfo{author}{\bibfnamefont{T.}~\bibnamefont{Opferkuch}},
  \bibnamefont{and}
  \bibinfo{author}{\bibfnamefont{P.}~\bibnamefont{Schwaller}},
  \bibinfo{journal}{JCAP} \textbf{\bibinfo{volume}{07}}, \bibinfo{pages}{007}
  (\bibinfo{year}{2019}), \eprint{1811.11175}.

\bibitem[{\citenamefont{Fairbairn et~al.}(2019)\citenamefont{Fairbairn, Hardy,
  and Wickens}}]{Fairbairn:2019xog}
\bibinfo{author}{\bibfnamefont{M.}~\bibnamefont{Fairbairn}},
  \bibinfo{author}{\bibfnamefont{E.}~\bibnamefont{Hardy}}, \bibnamefont{and}
  \bibinfo{author}{\bibfnamefont{A.}~\bibnamefont{Wickens}},
  \bibinfo{journal}{JHEP} \textbf{\bibinfo{volume}{07}}, \bibinfo{pages}{044}
  (\bibinfo{year}{2019}), \eprint{1901.11038}.

\bibitem[{\citenamefont{Baldes}(2017)}]{Baldes:2017rcu}
\bibinfo{author}{\bibfnamefont{I.}~\bibnamefont{Baldes}},
  \bibinfo{journal}{JCAP} \textbf{\bibinfo{volume}{05}}, \bibinfo{pages}{028}
  (\bibinfo{year}{2017}), \eprint{1702.02117}.

\bibitem[{\citenamefont{Tsumura et~al.}(2017)\citenamefont{Tsumura, Yamada, and
  Yamaguchi}}]{Tsumura:2017knk}
\bibinfo{author}{\bibfnamefont{K.}~\bibnamefont{Tsumura}},
  \bibinfo{author}{\bibfnamefont{M.}~\bibnamefont{Yamada}}, \bibnamefont{and}
  \bibinfo{author}{\bibfnamefont{Y.}~\bibnamefont{Yamaguchi}},
  \bibinfo{journal}{JCAP} \textbf{\bibinfo{volume}{07}}, \bibinfo{pages}{044}
  (\bibinfo{year}{2017}), \eprint{1704.00219}.

\bibitem[{\citenamefont{Aoki et~al.}(2017)\citenamefont{Aoki, Goto, and
  Kubo}}]{Aoki:2017aws}
\bibinfo{author}{\bibfnamefont{M.}~\bibnamefont{Aoki}},
  \bibinfo{author}{\bibfnamefont{H.}~\bibnamefont{Goto}}, \bibnamefont{and}
  \bibinfo{author}{\bibfnamefont{J.}~\bibnamefont{Kubo}},
  \bibinfo{journal}{Phys. Rev. D} \textbf{\bibinfo{volume}{96}},
  \bibinfo{pages}{075045} (\bibinfo{year}{2017}), \eprint{1709.07572}.

\bibitem[{\citenamefont{Croon and White}(2018)}]{Croon:2018new}
\bibinfo{author}{\bibfnamefont{D.}~\bibnamefont{Croon}} \bibnamefont{and}
  \bibinfo{author}{\bibfnamefont{G.}~\bibnamefont{White}},
  \bibinfo{journal}{JHEP} \textbf{\bibinfo{volume}{05}}, \bibinfo{pages}{210}
  (\bibinfo{year}{2018}), \eprint{1803.05438}.

\bibitem[{\citenamefont{Baldes and Garcia-Cely}(2019)}]{Baldes:2018emh}
\bibinfo{author}{\bibfnamefont{I.}~\bibnamefont{Baldes}} \bibnamefont{and}
  \bibinfo{author}{\bibfnamefont{C.}~\bibnamefont{Garcia-Cely}},
  \bibinfo{journal}{JHEP} \textbf{\bibinfo{volume}{05}}, \bibinfo{pages}{190}
  (\bibinfo{year}{2019}), \eprint{1809.01198}.

\bibitem[{\citenamefont{Foot et~al.}(2008)\citenamefont{Foot, Kobakhidze,
  McDonald, and Volkas}}]{Foot:2007iy}
\bibinfo{author}{\bibfnamefont{R.}~\bibnamefont{Foot}},
  \bibinfo{author}{\bibfnamefont{A.}~\bibnamefont{Kobakhidze}},
  \bibinfo{author}{\bibfnamefont{K.~L.} \bibnamefont{McDonald}},
  \bibnamefont{and} \bibinfo{author}{\bibfnamefont{R.~R.}
  \bibnamefont{Volkas}}, \bibinfo{journal}{Phys. Rev. D}
  \textbf{\bibinfo{volume}{77}}, \bibinfo{pages}{035006}
  (\bibinfo{year}{2008}), \eprint{0709.2750}.

\bibitem[{\citenamefont{Iso et~al.}(2009{\natexlab{a}})\citenamefont{Iso,
  Okada, and Orikasa}}]{Iso:2009nw}
\bibinfo{author}{\bibfnamefont{S.}~\bibnamefont{Iso}},
  \bibinfo{author}{\bibfnamefont{N.}~\bibnamefont{Okada}}, \bibnamefont{and}
  \bibinfo{author}{\bibfnamefont{Y.}~\bibnamefont{Orikasa}},
  \bibinfo{journal}{Phys. Rev. D} \textbf{\bibinfo{volume}{80}},
  \bibinfo{pages}{115007} (\bibinfo{year}{2009}{\natexlab{a}}),
  \eprint{0909.0128}.

\bibitem[{\citenamefont{Englert et~al.}(2013)\citenamefont{Englert, Jaeckel,
  Khoze, and Spannowsky}}]{Englert:2013gz}
\bibinfo{author}{\bibfnamefont{C.}~\bibnamefont{Englert}},
  \bibinfo{author}{\bibfnamefont{J.}~\bibnamefont{Jaeckel}},
  \bibinfo{author}{\bibfnamefont{V.~V.} \bibnamefont{Khoze}}, \bibnamefont{and}
  \bibinfo{author}{\bibfnamefont{M.}~\bibnamefont{Spannowsky}},
  \bibinfo{journal}{JHEP} \textbf{\bibinfo{volume}{04}}, \bibinfo{pages}{060}
  (\bibinfo{year}{2013}), \eprint{1301.4224}.

\bibitem[{\citenamefont{Farzinnia et~al.}(2013)\citenamefont{Farzinnia, He, and
  Ren}}]{Farzinnia:2013pga}
\bibinfo{author}{\bibfnamefont{A.}~\bibnamefont{Farzinnia}},
  \bibinfo{author}{\bibfnamefont{H.-J.} \bibnamefont{He}}, \bibnamefont{and}
  \bibinfo{author}{\bibfnamefont{J.}~\bibnamefont{Ren}},
  \bibinfo{journal}{Phys. Lett. B} \textbf{\bibinfo{volume}{727}},
  \bibinfo{pages}{141} (\bibinfo{year}{2013}), \eprint{1308.0295}.

\bibitem[{\citenamefont{Hur and Ko}(2011)}]{Hur:2011sv}
\bibinfo{author}{\bibfnamefont{T.}~\bibnamefont{Hur}} \bibnamefont{and}
  \bibinfo{author}{\bibfnamefont{P.}~\bibnamefont{Ko}}, \bibinfo{journal}{Phys.
  Rev. Lett.} \textbf{\bibinfo{volume}{106}}, \bibinfo{pages}{141802}
  (\bibinfo{year}{2011}), \eprint{1103.2571}.

\bibitem[{\citenamefont{Chang et~al.}(2007)\citenamefont{Chang, Ng, and
  Wu}}]{Chang:2007ki}
\bibinfo{author}{\bibfnamefont{W.-F.} \bibnamefont{Chang}},
  \bibinfo{author}{\bibfnamefont{J.~N.} \bibnamefont{Ng}}, \bibnamefont{and}
  \bibinfo{author}{\bibfnamefont{J.~M.~S.} \bibnamefont{Wu}},
  \bibinfo{journal}{Phys. Rev. D} \textbf{\bibinfo{volume}{75}},
  \bibinfo{pages}{115016} (\bibinfo{year}{2007}), \eprint{hep-ph/0701254}.

\bibitem[{\citenamefont{Iso et~al.}(2009{\natexlab{b}})\citenamefont{Iso,
  Okada, and Orikasa}}]{Iso:2009ss}
\bibinfo{author}{\bibfnamefont{S.}~\bibnamefont{Iso}},
  \bibinfo{author}{\bibfnamefont{N.}~\bibnamefont{Okada}}, \bibnamefont{and}
  \bibinfo{author}{\bibfnamefont{Y.}~\bibnamefont{Orikasa}},
  \bibinfo{journal}{Phys. Lett. B} \textbf{\bibinfo{volume}{676}},
  \bibinfo{pages}{81} (\bibinfo{year}{2009}{\natexlab{b}}), \eprint{0902.4050}.

\bibitem[{\citenamefont{Vilenkin and Vachaspati}(1987)}]{Vilenkin:1986ku}
\bibinfo{author}{\bibfnamefont{A.}~\bibnamefont{Vilenkin}} \bibnamefont{and}
  \bibinfo{author}{\bibfnamefont{T.}~\bibnamefont{Vachaspati}},
  \bibinfo{journal}{Phys. Rev. D} \textbf{\bibinfo{volume}{35}},
  \bibinfo{pages}{1138} (\bibinfo{year}{1987}).

\bibitem[{\citenamefont{Davis}(1986)}]{Davis:1986xc}
\bibinfo{author}{\bibfnamefont{R.~L.} \bibnamefont{Davis}},
  \bibinfo{journal}{Phys. Lett. B} \textbf{\bibinfo{volume}{180}},
  \bibinfo{pages}{225} (\bibinfo{year}{1986}).

\bibitem[{\citenamefont{Harari and Sikivie}(1987)}]{Harari:1987ht}
\bibinfo{author}{\bibfnamefont{D.}~\bibnamefont{Harari}} \bibnamefont{and}
  \bibinfo{author}{\bibfnamefont{P.}~\bibnamefont{Sikivie}},
  \bibinfo{journal}{Phys. Lett. B} \textbf{\bibinfo{volume}{195}},
  \bibinfo{pages}{361} (\bibinfo{year}{1987}).

\bibitem[{\citenamefont{Hagmann and Sikivie}(1991)}]{Hagmann:1990mj}
\bibinfo{author}{\bibfnamefont{C.}~\bibnamefont{Hagmann}} \bibnamefont{and}
  \bibinfo{author}{\bibfnamefont{P.}~\bibnamefont{Sikivie}},
  \bibinfo{journal}{Nucl. Phys. B} \textbf{\bibinfo{volume}{363}},
  \bibinfo{pages}{247} (\bibinfo{year}{1991}).

\bibitem[{\citenamefont{Battye and
  Shellard}(1994{\natexlab{a}})}]{Battye:1993jv}
\bibinfo{author}{\bibfnamefont{R.~A.} \bibnamefont{Battye}} \bibnamefont{and}
  \bibinfo{author}{\bibfnamefont{E.~P.~S.} \bibnamefont{Shellard}},
  \bibinfo{journal}{Nucl. Phys. B} \textbf{\bibinfo{volume}{423}},
  \bibinfo{pages}{260} (\bibinfo{year}{1994}{\natexlab{a}}),
  \eprint{astro-ph/9311017}.

\bibitem[{\citenamefont{Battye and
  Shellard}(1994{\natexlab{b}})}]{Battye:1994au}
\bibinfo{author}{\bibfnamefont{R.~A.} \bibnamefont{Battye}} \bibnamefont{and}
  \bibinfo{author}{\bibfnamefont{E.~P.~S.} \bibnamefont{Shellard}},
  \bibinfo{journal}{Phys. Rev. Lett.} \textbf{\bibinfo{volume}{73}},
  \bibinfo{pages}{2954} (\bibinfo{year}{1994}{\natexlab{b}}),
  \bibinfo{note}{[Erratum: Phys.Rev.Lett. 76, 2203--2204 (1996)]},
  \eprint{astro-ph/9403018}.

\bibitem[{\citenamefont{Yamaguchi et~al.}(1999)\citenamefont{Yamaguchi,
  Kawasaki, and Yokoyama}}]{Yamaguchi:1998gx}
\bibinfo{author}{\bibfnamefont{M.}~\bibnamefont{Yamaguchi}},
  \bibinfo{author}{\bibfnamefont{M.}~\bibnamefont{Kawasaki}}, \bibnamefont{and}
  \bibinfo{author}{\bibfnamefont{J.}~\bibnamefont{Yokoyama}},
  \bibinfo{journal}{Phys. Rev. Lett.} \textbf{\bibinfo{volume}{82}},
  \bibinfo{pages}{4578} (\bibinfo{year}{1999}), \eprint{hep-ph/9811311}.

\bibitem[{\citenamefont{Hagmann et~al.}(2001)\citenamefont{Hagmann, Chang, and
  Sikivie}}]{Hagmann:2000ja}
\bibinfo{author}{\bibfnamefont{C.}~\bibnamefont{Hagmann}},
  \bibinfo{author}{\bibfnamefont{S.}~\bibnamefont{Chang}}, \bibnamefont{and}
  \bibinfo{author}{\bibfnamefont{P.}~\bibnamefont{Sikivie}},
  \bibinfo{journal}{Phys. Rev. D} \textbf{\bibinfo{volume}{63}},
  \bibinfo{pages}{125018} (\bibinfo{year}{2001}), \eprint{hep-ph/0012361}.

\bibitem[{\citenamefont{Kibble}(1976)}]{Kibble:1976sj}
\bibinfo{author}{\bibfnamefont{T.~W.~B.} \bibnamefont{Kibble}},
  \bibinfo{journal}{J. Phys. A} \textbf{\bibinfo{volume}{9}},
  \bibinfo{pages}{1387} (\bibinfo{year}{1976}).

\bibitem[{\citenamefont{Hindmarsh and Kibble}(1995)}]{Hindmarsh:1994re}
\bibinfo{author}{\bibfnamefont{M.~B.} \bibnamefont{Hindmarsh}}
  \bibnamefont{and} \bibinfo{author}{\bibfnamefont{T.~W.~B.}
  \bibnamefont{Kibble}}, \bibinfo{journal}{Rept. Prog. Phys.}
  \textbf{\bibinfo{volume}{58}}, \bibinfo{pages}{477} (\bibinfo{year}{1995}),
  \eprint{hep-ph/9411342}.

\bibitem[{\citenamefont{Dev et~al.}(2019)\citenamefont{Dev, Ferrer, Zhang, and
  Zhang}}]{Dev:2019njv}
\bibinfo{author}{\bibfnamefont{P.~S.~B.} \bibnamefont{Dev}},
  \bibinfo{author}{\bibfnamefont{F.}~\bibnamefont{Ferrer}},
  \bibinfo{author}{\bibfnamefont{Y.}~\bibnamefont{Zhang}}, \bibnamefont{and}
  \bibinfo{author}{\bibfnamefont{Y.}~\bibnamefont{Zhang}},
  \bibinfo{journal}{JCAP} \textbf{\bibinfo{volume}{11}}, \bibinfo{pages}{006}
  (\bibinfo{year}{2019}), \eprint{1905.00891}.

\bibitem[{\citenamefont{Von~Harling et~al.}(2020)\citenamefont{Von~Harling,
  Pomarol, Pujol\`as, and Rompineve}}]{VonHarling:2019rgb}
\bibinfo{author}{\bibfnamefont{B.}~\bibnamefont{Von~Harling}},
  \bibinfo{author}{\bibfnamefont{A.}~\bibnamefont{Pomarol}},
  \bibinfo{author}{\bibfnamefont{O.}~\bibnamefont{Pujol\`as}},
  \bibnamefont{and}
  \bibinfo{author}{\bibfnamefont{F.}~\bibnamefont{Rompineve}},
  \bibinfo{journal}{JHEP} \textbf{\bibinfo{volume}{04}}, \bibinfo{pages}{195}
  (\bibinfo{year}{2020}), \eprint{1912.07587}.

\bibitem[{\citenamefont{Ghoshal and Salvio}(2020)}]{Ghoshal:2020vud}
\bibinfo{author}{\bibfnamefont{A.}~\bibnamefont{Ghoshal}} \bibnamefont{and}
  \bibinfo{author}{\bibfnamefont{A.}~\bibnamefont{Salvio}},
  \bibinfo{journal}{JHEP} \textbf{\bibinfo{volume}{12}}, \bibinfo{pages}{049}
  (\bibinfo{year}{2020}), \eprint{2007.00005}.

\bibitem[{\citenamefont{Delle~Rose et~al.}(2020)\citenamefont{Delle~Rose,
  Panico, Redi, and Tesi}}]{DelleRose:2019pgi}
\bibinfo{author}{\bibfnamefont{L.}~\bibnamefont{Delle~Rose}},
  \bibinfo{author}{\bibfnamefont{G.}~\bibnamefont{Panico}},
  \bibinfo{author}{\bibfnamefont{M.}~\bibnamefont{Redi}}, \bibnamefont{and}
  \bibinfo{author}{\bibfnamefont{A.}~\bibnamefont{Tesi}},
  \bibinfo{journal}{JHEP} \textbf{\bibinfo{volume}{04}}, \bibinfo{pages}{025}
  (\bibinfo{year}{2020}), \eprint{1912.06139}.

\bibitem[{\citenamefont{Giblin and Mertens}(2014)}]{Giblin:2014qia}
\bibinfo{author}{\bibfnamefont{J.~T.} \bibnamefont{Giblin}} \bibnamefont{and}
  \bibinfo{author}{\bibfnamefont{J.~B.} \bibnamefont{Mertens}},
  \bibinfo{journal}{Phys. Rev. D} \textbf{\bibinfo{volume}{90}},
  \bibinfo{pages}{023532} (\bibinfo{year}{2014}), \eprint{1405.4005}.

\bibitem[{\citenamefont{Hindmarsh et~al.}(2014)\citenamefont{Hindmarsh, Huber,
  Rummukainen, and Weir}}]{Hindmarsh:2013xza}
\bibinfo{author}{\bibfnamefont{M.}~\bibnamefont{Hindmarsh}},
  \bibinfo{author}{\bibfnamefont{S.~J.} \bibnamefont{Huber}},
  \bibinfo{author}{\bibfnamefont{K.}~\bibnamefont{Rummukainen}},
  \bibnamefont{and} \bibinfo{author}{\bibfnamefont{D.~J.} \bibnamefont{Weir}},
  \bibinfo{journal}{Phys. Rev. Lett.} \textbf{\bibinfo{volume}{112}},
  \bibinfo{pages}{041301} (\bibinfo{year}{2014}), \eprint{1304.2433}.

\bibitem[{\citenamefont{Cutting
  et~al.}(2020{\natexlab{a}})\citenamefont{Cutting, Hindmarsh, and
  Weir}}]{Cutting:2019zws}
\bibinfo{author}{\bibfnamefont{D.}~\bibnamefont{Cutting}},
  \bibinfo{author}{\bibfnamefont{M.}~\bibnamefont{Hindmarsh}},
  \bibnamefont{and} \bibinfo{author}{\bibfnamefont{D.~J.} \bibnamefont{Weir}},
  \bibinfo{journal}{Phys. Rev. Lett.} \textbf{\bibinfo{volume}{125}},
  \bibinfo{pages}{021302} (\bibinfo{year}{2020}{\natexlab{a}}),
  \eprint{1906.00480}.

\bibitem[{\citenamefont{Hindmarsh et~al.}(2015)\citenamefont{Hindmarsh, Huber,
  Rummukainen, and Weir}}]{Hindmarsh:2015qta}
\bibinfo{author}{\bibfnamefont{M.}~\bibnamefont{Hindmarsh}},
  \bibinfo{author}{\bibfnamefont{S.~J.} \bibnamefont{Huber}},
  \bibinfo{author}{\bibfnamefont{K.}~\bibnamefont{Rummukainen}},
  \bibnamefont{and} \bibinfo{author}{\bibfnamefont{D.~J.} \bibnamefont{Weir}},
  \bibinfo{journal}{Phys. Rev. D} \textbf{\bibinfo{volume}{92}},
  \bibinfo{pages}{123009} (\bibinfo{year}{2015}), \eprint{1504.03291}.

\bibitem[{\citenamefont{Hindmarsh et~al.}(2017)\citenamefont{Hindmarsh, Huber,
  Rummukainen, and Weir}}]{Hindmarsh:2017gnf}
\bibinfo{author}{\bibfnamefont{M.}~\bibnamefont{Hindmarsh}},
  \bibinfo{author}{\bibfnamefont{S.~J.} \bibnamefont{Huber}},
  \bibinfo{author}{\bibfnamefont{K.}~\bibnamefont{Rummukainen}},
  \bibnamefont{and} \bibinfo{author}{\bibfnamefont{D.~J.} \bibnamefont{Weir}},
  \bibinfo{journal}{Phys. Rev. D} \textbf{\bibinfo{volume}{96}},
  \bibinfo{pages}{103520} (\bibinfo{year}{2017}), \bibinfo{note}{[Erratum:
  Phys.Rev.D 101, 089902 (2020)]}, \eprint{1704.05871}.

\bibitem[{\citenamefont{Cutting et~al.}(2018)\citenamefont{Cutting, Hindmarsh,
  and Weir}}]{Cutting:2018tjt}
\bibinfo{author}{\bibfnamefont{D.}~\bibnamefont{Cutting}},
  \bibinfo{author}{\bibfnamefont{M.}~\bibnamefont{Hindmarsh}},
  \bibnamefont{and} \bibinfo{author}{\bibfnamefont{D.~J.} \bibnamefont{Weir}},
  \bibinfo{journal}{Phys. Rev. D} \textbf{\bibinfo{volume}{97}},
  \bibinfo{pages}{123513} (\bibinfo{year}{2018}), \eprint{1802.05712}.

\bibitem[{\citenamefont{Cutting
  et~al.}(2020{\natexlab{b}})\citenamefont{Cutting, Escartin, Hindmarsh, and
  Weir}}]{Cutting:2020nla}
\bibinfo{author}{\bibfnamefont{D.}~\bibnamefont{Cutting}},
  \bibinfo{author}{\bibfnamefont{E.~G.} \bibnamefont{Escartin}},
  \bibinfo{author}{\bibfnamefont{M.}~\bibnamefont{Hindmarsh}},
  \bibnamefont{and} \bibinfo{author}{\bibfnamefont{D.~J.} \bibnamefont{Weir}}
  (\bibinfo{year}{2020}{\natexlab{b}}), \eprint{2005.13537}.

\bibitem[{\citenamefont{Roper~Pol et~al.}(2020)\citenamefont{Roper~Pol, Mandal,
  Brandenburg, Kahniashvili, and Kosowsky}}]{Pol:2019yex}
\bibinfo{author}{\bibfnamefont{A.}~\bibnamefont{Roper~Pol}},
  \bibinfo{author}{\bibfnamefont{S.}~\bibnamefont{Mandal}},
  \bibinfo{author}{\bibfnamefont{A.}~\bibnamefont{Brandenburg}},
  \bibinfo{author}{\bibfnamefont{T.}~\bibnamefont{Kahniashvili}},
  \bibnamefont{and} \bibinfo{author}{\bibfnamefont{A.}~\bibnamefont{Kosowsky}},
  \bibinfo{journal}{Phys. Rev. D} \textbf{\bibinfo{volume}{102}},
  \bibinfo{pages}{083512} (\bibinfo{year}{2020}), \eprint{1903.08585}.

\bibitem[{\citenamefont{Jinno and Takimoto}(2017)}]{Jinno:2016knw}
\bibinfo{author}{\bibfnamefont{R.}~\bibnamefont{Jinno}} \bibnamefont{and}
  \bibinfo{author}{\bibfnamefont{M.}~\bibnamefont{Takimoto}},
  \bibinfo{journal}{Phys. Rev. D} \textbf{\bibinfo{volume}{95}},
  \bibinfo{pages}{015020} (\bibinfo{year}{2017}), \eprint{1604.05035}.

\bibitem[{\citenamefont{Marzola et~al.}(2017)\citenamefont{Marzola, Racioppi,
  and Vaskonen}}]{Marzola:2017jzl}
\bibinfo{author}{\bibfnamefont{L.}~\bibnamefont{Marzola}},
  \bibinfo{author}{\bibfnamefont{A.}~\bibnamefont{Racioppi}}, \bibnamefont{and}
  \bibinfo{author}{\bibfnamefont{V.}~\bibnamefont{Vaskonen}},
  \bibinfo{journal}{Eur. Phys. J. C} \textbf{\bibinfo{volume}{77}},
  \bibinfo{pages}{484} (\bibinfo{year}{2017}), \eprint{1704.01034}.

\bibitem[{\citenamefont{Iso et~al.}(2017)\citenamefont{Iso, Serpico, and
  Shimada}}]{Iso:2017uuu}
\bibinfo{author}{\bibfnamefont{S.}~\bibnamefont{Iso}},
  \bibinfo{author}{\bibfnamefont{P.~D.} \bibnamefont{Serpico}},
  \bibnamefont{and} \bibinfo{author}{\bibfnamefont{K.}~\bibnamefont{Shimada}},
  \bibinfo{journal}{Phys. Rev. Lett.} \textbf{\bibinfo{volume}{119}},
  \bibinfo{pages}{141301} (\bibinfo{year}{2017}), \eprint{1704.04955}.

\bibitem[{\citenamefont{Lewicki and
  Vaskonen}(2020{\natexlab{a}})}]{Lewicki:2020jiv}
\bibinfo{author}{\bibfnamefont{M.}~\bibnamefont{Lewicki}} \bibnamefont{and}
  \bibinfo{author}{\bibfnamefont{V.}~\bibnamefont{Vaskonen}},
  \bibinfo{journal}{Eur. Phys. J. C} \textbf{\bibinfo{volume}{80}},
  \bibinfo{pages}{1003} (\bibinfo{year}{2020}{\natexlab{a}}),
  \eprint{2007.04967}.

\bibitem[{\citenamefont{Buschmann et~al.}(2020)\citenamefont{Buschmann, Foster,
  and Safdi}}]{Buschmann:2019icd}
\bibinfo{author}{\bibfnamefont{M.}~\bibnamefont{Buschmann}},
  \bibinfo{author}{\bibfnamefont{J.~W.} \bibnamefont{Foster}},
  \bibnamefont{and} \bibinfo{author}{\bibfnamefont{B.~R.} \bibnamefont{Safdi}},
  \bibinfo{journal}{Phys. Rev. Lett.} \textbf{\bibinfo{volume}{124}},
  \bibinfo{pages}{161103} (\bibinfo{year}{2020}), \eprint{1906.00967}.

\bibitem[{\citenamefont{Gorghetto et~al.}(2018)\citenamefont{Gorghetto, Hardy,
  and Villadoro}}]{Gorghetto:2018myk}
\bibinfo{author}{\bibfnamefont{M.}~\bibnamefont{Gorghetto}},
  \bibinfo{author}{\bibfnamefont{E.}~\bibnamefont{Hardy}}, \bibnamefont{and}
  \bibinfo{author}{\bibfnamefont{G.}~\bibnamefont{Villadoro}},
  \bibinfo{journal}{JHEP} \textbf{\bibinfo{volume}{07}}, \bibinfo{pages}{151}
  (\bibinfo{year}{2018}), \eprint{1806.04677}.

\bibitem[{\citenamefont{Figueroa et~al.}(2020)\citenamefont{Figueroa,
  Hindmarsh, Lizarraga, and Urrestilla}}]{Figueroa:2020lvo}
\bibinfo{author}{\bibfnamefont{D.~G.} \bibnamefont{Figueroa}},
  \bibinfo{author}{\bibfnamefont{M.}~\bibnamefont{Hindmarsh}},
  \bibinfo{author}{\bibfnamefont{J.}~\bibnamefont{Lizarraga}},
  \bibnamefont{and}
  \bibinfo{author}{\bibfnamefont{J.}~\bibnamefont{Urrestilla}},
  \bibinfo{journal}{Phys. Rev. D} \textbf{\bibinfo{volume}{102}},
  \bibinfo{pages}{103516} (\bibinfo{year}{2020}), \eprint{2007.03337}.

\bibitem[{\citenamefont{Gorghetto et~al.}(2021)\citenamefont{Gorghetto, Hardy,
  and Nicolaescu}}]{Gorghetto:2021fsn}
\bibinfo{author}{\bibfnamefont{M.}~\bibnamefont{Gorghetto}},
  \bibinfo{author}{\bibfnamefont{E.}~\bibnamefont{Hardy}}, \bibnamefont{and}
  \bibinfo{author}{\bibfnamefont{H.}~\bibnamefont{Nicolaescu}},
  \bibinfo{journal}{JCAP} \textbf{\bibinfo{volume}{06}}, \bibinfo{pages}{034}
  (\bibinfo{year}{2021}), \eprint{2101.11007}.

\bibitem[{\citenamefont{Chang and Cui}(2022)}]{Chang:2021afa}
\bibinfo{author}{\bibfnamefont{C.-F.} \bibnamefont{Chang}} \bibnamefont{and}
  \bibinfo{author}{\bibfnamefont{Y.}~\bibnamefont{Cui}},
  \bibinfo{journal}{JHEP} \textbf{\bibinfo{volume}{03}}, \bibinfo{pages}{114}
  (\bibinfo{year}{2022}), \eprint{2106.09746}.

\bibitem[{\citenamefont{Borrill et~al.}(1995)\citenamefont{Borrill, Kibble,
  Vachaspati, and Vilenkin}}]{Borrill:1995gu}
\bibinfo{author}{\bibfnamefont{J.}~\bibnamefont{Borrill}},
  \bibinfo{author}{\bibfnamefont{T.~W.~B.} \bibnamefont{Kibble}},
  \bibinfo{author}{\bibfnamefont{T.}~\bibnamefont{Vachaspati}},
  \bibnamefont{and} \bibinfo{author}{\bibfnamefont{A.}~\bibnamefont{Vilenkin}},
  \bibinfo{journal}{Phys. Rev. D} \textbf{\bibinfo{volume}{52}},
  \bibinfo{pages}{1934} (\bibinfo{year}{1995}), \eprint{hep-ph/9503223}.

\bibitem[{\citenamefont{Digal et~al.}(1997)\citenamefont{Digal, Sengupta, and
  Srivastava}}]{Digal:1997ip}
\bibinfo{author}{\bibfnamefont{S.}~\bibnamefont{Digal}},
  \bibinfo{author}{\bibfnamefont{S.}~\bibnamefont{Sengupta}}, \bibnamefont{and}
  \bibinfo{author}{\bibfnamefont{A.~M.} \bibnamefont{Srivastava}},
  \bibinfo{journal}{Phys. Rev. D} \textbf{\bibinfo{volume}{56}},
  \bibinfo{pages}{2035} (\bibinfo{year}{1997}), \eprint{hep-ph/9705246}.

\bibitem[{\citenamefont{Copeland and Saffin}(1996)}]{Copeland:1996jz}
\bibinfo{author}{\bibfnamefont{E.~J.} \bibnamefont{Copeland}} \bibnamefont{and}
  \bibinfo{author}{\bibfnamefont{P.~M.} \bibnamefont{Saffin}},
  \bibinfo{journal}{Phys. Rev. D} \textbf{\bibinfo{volume}{54}},
  \bibinfo{pages}{6088} (\bibinfo{year}{1996}), \eprint{hep-ph/9604231}.

\bibitem[{\citenamefont{Digal et~al.}(1998)\citenamefont{Digal, Sengupta, and
  Srivastava}}]{Digal:1997gc}
\bibinfo{author}{\bibfnamefont{S.}~\bibnamefont{Digal}},
  \bibinfo{author}{\bibfnamefont{S.}~\bibnamefont{Sengupta}}, \bibnamefont{and}
  \bibinfo{author}{\bibfnamefont{A.~M.} \bibnamefont{Srivastava}},
  \bibinfo{journal}{Phys. Rev. D} \textbf{\bibinfo{volume}{58}},
  \bibinfo{pages}{103510} (\bibinfo{year}{1998}), \eprint{hep-ph/9707221}.

\bibitem[{\citenamefont{Ferrera and Melfo}(1996)}]{Ferrera:1995ef}
\bibinfo{author}{\bibfnamefont{A.}~\bibnamefont{Ferrera}} \bibnamefont{and}
  \bibinfo{author}{\bibfnamefont{A.}~\bibnamefont{Melfo}},
  \bibinfo{journal}{Phys. Rev. D} \textbf{\bibinfo{volume}{53}},
  \bibinfo{pages}{6852} (\bibinfo{year}{1996}), \eprint{hep-ph/9512290}.

\bibitem[{\citenamefont{Ferrera}(1999)}]{Ferrera:1997xg}
\bibinfo{author}{\bibfnamefont{A.}~\bibnamefont{Ferrera}},
  \bibinfo{journal}{Phys. Rev. D} \textbf{\bibinfo{volume}{59}},
  \bibinfo{pages}{123503} (\bibinfo{year}{1999}), \eprint{hep-ph/9811269}.

\bibitem[{\citenamefont{Ferrera}(1998)}]{Ferrera:1996hu}
\bibinfo{author}{\bibfnamefont{A.}~\bibnamefont{Ferrera}},
  \bibinfo{journal}{Phys. Rev. D} \textbf{\bibinfo{volume}{57}},
  \bibinfo{pages}{7130} (\bibinfo{year}{1998}), \eprint{hep-ph/9612487}.

\bibitem[{\citenamefont{Copeland et~al.}(2000)\citenamefont{Copeland, Saffin,
  and Tornkvist}}]{Copeland:1999ua}
\bibinfo{author}{\bibfnamefont{E.~J.} \bibnamefont{Copeland}},
  \bibinfo{author}{\bibfnamefont{P.~M.} \bibnamefont{Saffin}},
  \bibnamefont{and}
  \bibinfo{author}{\bibfnamefont{O.}~\bibnamefont{Tornkvist}},
  \bibinfo{journal}{Phys. Rev. D} \textbf{\bibinfo{volume}{61}},
  \bibinfo{pages}{105005} (\bibinfo{year}{2000}), \eprint{hep-ph/9907437}.

\bibitem[{\citenamefont{Davis and Lilley}(2000)}]{Davis:1999ii}
\bibinfo{author}{\bibfnamefont{A.-C.} \bibnamefont{Davis}} \bibnamefont{and}
  \bibinfo{author}{\bibfnamefont{M.}~\bibnamefont{Lilley}},
  \bibinfo{journal}{Phys. Rev. D} \textbf{\bibinfo{volume}{61}},
  \bibinfo{pages}{043502} (\bibinfo{year}{2000}), \eprint{hep-ph/9908398}.

\bibitem[{\citenamefont{Lilley and Ferrera}(2001)}]{Lilley:2001df}
\bibinfo{author}{\bibfnamefont{M.}~\bibnamefont{Lilley}} \bibnamefont{and}
  \bibinfo{author}{\bibfnamefont{A.}~\bibnamefont{Ferrera}},
  \bibinfo{journal}{Phys. Rev. D} \textbf{\bibinfo{volume}{64}},
  \bibinfo{pages}{023520} (\bibinfo{year}{2001}), \eprint{hep-ph/0102035}.

\bibitem[{\citenamefont{Masoumi et~al.}(2017)\citenamefont{Masoumi, Olum, and
  Wachter}}]{Masoumi:2017trx}
\bibinfo{author}{\bibfnamefont{A.}~\bibnamefont{Masoumi}},
  \bibinfo{author}{\bibfnamefont{K.~D.} \bibnamefont{Olum}}, \bibnamefont{and}
  \bibinfo{author}{\bibfnamefont{J.~M.} \bibnamefont{Wachter}},
  \bibinfo{journal}{JCAP} \textbf{\bibinfo{volume}{10}}, \bibinfo{pages}{022}
  (\bibinfo{year}{2017}), \eprint{1702.00356}.

\bibitem[{\citenamefont{Guada et~al.}(2020)\citenamefont{Guada, Nemev\v{s}ek,
  and Pintar}}]{Guada:2020xnz}
\bibinfo{author}{\bibfnamefont{V.}~\bibnamefont{Guada}},
  \bibinfo{author}{\bibfnamefont{M.}~\bibnamefont{Nemev\v{s}ek}},
  \bibnamefont{and} \bibinfo{author}{\bibfnamefont{M.}~\bibnamefont{Pintar}},
  \bibinfo{journal}{Comput. Phys. Commun.} \textbf{\bibinfo{volume}{256}},
  \bibinfo{pages}{107480} (\bibinfo{year}{2020}), \eprint{2002.00881}.

\bibitem[{\citenamefont{Enqvist et~al.}(1992)\citenamefont{Enqvist, Ignatius,
  Kajantie, and Rummukainen}}]{Enqvist:1991xw}
\bibinfo{author}{\bibfnamefont{K.}~\bibnamefont{Enqvist}},
  \bibinfo{author}{\bibfnamefont{J.}~\bibnamefont{Ignatius}},
  \bibinfo{author}{\bibfnamefont{K.}~\bibnamefont{Kajantie}}, \bibnamefont{and}
  \bibinfo{author}{\bibfnamefont{K.}~\bibnamefont{Rummukainen}},
  \bibinfo{journal}{Phys. Rev. D} \textbf{\bibinfo{volume}{45}},
  \bibinfo{pages}{3415} (\bibinfo{year}{1992}).

\bibitem[{\citenamefont{Konstandin}(2018)}]{Konstandin:2017sat}
\bibinfo{author}{\bibfnamefont{T.}~\bibnamefont{Konstandin}},
  \bibinfo{journal}{JCAP} \textbf{\bibinfo{volume}{03}}, \bibinfo{pages}{047}
  (\bibinfo{year}{2018}), \eprint{1712.06869}.

\bibitem[{\citenamefont{Ellis et~al.}(2020)\citenamefont{Ellis, Lewicki, and
  Vaskonen}}]{Ellis:2020nnr}
\bibinfo{author}{\bibfnamefont{J.}~\bibnamefont{Ellis}},
  \bibinfo{author}{\bibfnamefont{M.}~\bibnamefont{Lewicki}}, \bibnamefont{and}
  \bibinfo{author}{\bibfnamefont{V.}~\bibnamefont{Vaskonen}},
  \bibinfo{journal}{JCAP} \textbf{\bibinfo{volume}{11}}, \bibinfo{pages}{020}
  (\bibinfo{year}{2020}), \eprint{2007.15586}.

\bibitem[{\citenamefont{Lewicki and
  Vaskonen}(2020{\natexlab{b}})}]{Lewicki:2020azd}
\bibinfo{author}{\bibfnamefont{M.}~\bibnamefont{Lewicki}} \bibnamefont{and}
  \bibinfo{author}{\bibfnamefont{V.}~\bibnamefont{Vaskonen}}
  (\bibinfo{year}{2020}{\natexlab{b}}), \eprint{2012.07826}.

\bibitem[{\citenamefont{Garcia-Bellido
  et~al.}(2008)\citenamefont{Garcia-Bellido, Figueroa, and
  Sastre}}]{GarciaBellido:2007af}
\bibinfo{author}{\bibfnamefont{J.}~\bibnamefont{Garcia-Bellido}},
  \bibinfo{author}{\bibfnamefont{D.~G.} \bibnamefont{Figueroa}},
  \bibnamefont{and} \bibinfo{author}{\bibfnamefont{A.}~\bibnamefont{Sastre}},
  \bibinfo{journal}{Phys. Rev. D} \textbf{\bibinfo{volume}{77}},
  \bibinfo{pages}{043517} (\bibinfo{year}{2008}), \eprint{0707.0839}.

\bibitem[{\citenamefont{Adshead et~al.}(2020)\citenamefont{Adshead, Giblin,
  Pieroni, and Weiner}}]{Adshead:2019lbr}
\bibinfo{author}{\bibfnamefont{P.}~\bibnamefont{Adshead}},
  \bibinfo{author}{\bibfnamefont{J.~T.} \bibnamefont{Giblin}},
  \bibinfo{author}{\bibfnamefont{M.}~\bibnamefont{Pieroni}}, \bibnamefont{and}
  \bibinfo{author}{\bibfnamefont{Z.~J.} \bibnamefont{Weiner}},
  \bibinfo{journal}{Phys. Rev. D} \textbf{\bibinfo{volume}{101}},
  \bibinfo{pages}{083534} (\bibinfo{year}{2020}), \eprint{1909.12842}.

\bibitem[{\citenamefont{Hindmarsh and Hijazi}(2019)}]{Hindmarsh:2019phv}
\bibinfo{author}{\bibfnamefont{M.}~\bibnamefont{Hindmarsh}} \bibnamefont{and}
  \bibinfo{author}{\bibfnamefont{M.}~\bibnamefont{Hijazi}},
  \bibinfo{journal}{JCAP} \textbf{\bibinfo{volume}{1912}}, \bibinfo{pages}{062}
  (\bibinfo{year}{2019}), \eprint{1909.10040}.

\bibitem[{\citenamefont{Saurabh et~al.}(2020)\citenamefont{Saurabh, Vachaspati,
  and Pogosian}}]{Saurabh:2020pqe}
\bibinfo{author}{\bibfnamefont{A.}~\bibnamefont{Saurabh}},
  \bibinfo{author}{\bibfnamefont{T.}~\bibnamefont{Vachaspati}},
  \bibnamefont{and} \bibinfo{author}{\bibfnamefont{L.}~\bibnamefont{Pogosian}},
  \bibinfo{journal}{Phys. Rev. D} \textbf{\bibinfo{volume}{101}},
  \bibinfo{pages}{083522} (\bibinfo{year}{2020}), \eprint{2001.01030}.

\bibitem[{\citenamefont{Vilenkin and Everett}(1982)}]{Vilenkin:1982ks}
\bibinfo{author}{\bibfnamefont{A.}~\bibnamefont{Vilenkin}} \bibnamefont{and}
  \bibinfo{author}{\bibfnamefont{A.~E.} \bibnamefont{Everett}},
  \bibinfo{journal}{Phys. Rev. Lett.} \textbf{\bibinfo{volume}{48}},
  \bibinfo{pages}{1867} (\bibinfo{year}{1982}).

\end{thebibliography}

\end{document}